\documentclass[11pt,a4paper]{article}
\usepackage{jcappub}

\usepackage{graphicx}
\usepackage{txfonts}
\usepackage{lscape}
\usepackage[left]{lineno}

\usepackage{threeparttable}		

\usepackage{natbib}

\notoc

\begin{document}

\title{Optical and X-ray early follow-up of ANTARES neutrino alerts}

\author[1]{S.~Adri\'an-Mart\'inez,}
\author[5]{M.~Ageron,}
\author[2]{A.~Albert,}
\author[5,a]{I.~Al Samarai,%
\note[a]{Now at Laboratoire de Physique Nucl\'eaire et de Hautes Energies (LPNHE), Universit\'es Paris 6 et Paris 7, CNRS-IN2P3, Paris, France}}
\author[3]{M.~Andr\'e,}
\author[4]{G.~Anton,}
\author[1]{M.~Ardid,}
\author[5]{J.-J.~Aubert,}
\author[6]{B.~Baret,}
\author[7]{J.~Barrios-Mart\'{\i},}
% ...
\author{(the ANTARES Collaboration),} 
\author[38,39]{A.~Klotz,}
\author[40]{M.~Boer,} 
\author[41]{A.~Le Van Suu}
\author{(members of the TAROT Collaboration),} 
\author[42]{C.~Akerlof,}
\author[43]{W.~Zheng}
\author{(members of the ROTSE Collaboration),}
\author[44]{P.~Evans,}
\author[45]{N.~Gehrels,}
\author[46]{J.~Kennea,}
\author[44]{J.P.~Osborne}
\author{(members of the \textit{Swift} Collaboration),}
\author[47]{D.M.~Coward}
\author{(member of the Zadko Collaboration)}

\affiliation[1]{\small{Institut d'Investigaci\'o per a la Gesti\'o Integrada de les Zones Costaneres (IGIC) - Universitat Polit\`ecnica de Val\`encia. C/  Paranimf 1 , 46730 Gandia, Spain}}
\affiliation[2]{\small{GRPHE - Universit\'e de Haute Alsace - Institut universitaire de technologie de Colmar, 34 rue du Grillenbreit BP 50568 - 68008 Colmar, France}}
\affiliation[3]{\small{Technical University of Catalonia, Laboratory of Applied Bioacoustics, Rambla Exposici\'o,08800 Vilanova i la Geltr\'u,Barcelona, Spain}}
\affiliation[4]{\small{Friedrich-Alexander-Universit\"at Erlangen-N\"urnberg, Erlangen Centre for Astroparticle Physics, Erwin-Rommel-Str. 1, 91058 Erlangen, Germany}}
\affiliation[5]{\small{Aix Marseille Universit\'e, CNRS/IN2P3, CPPM UMR 7346, 13288, Marseille, France}}
\affiliation[6]{\small{APC, Universit\'e Paris Diderot, CNRS/IN2P3, CEA/IRFU, Observatoire de Paris, Sorbonne Paris Cit\'e, 75205 Paris, France}}
\affiliation[7]{\small{IFIC - Instituto de F\'isica Corpuscular, Edificios Investigaci\'on de Paterna, CSIC - Universitat de Val\`encia, Apdo. de Correos 22085, 46071 Valencia, Spain}}
% ...
\affiliation[38]{\small{Universit\'e de Toulouse; UPS-OMP; IRAP; Toulouse, France}}
\affiliation[39]{\small{CNRS; IRAP; 14, avenue Eduard-Belin, F-31400 Toulouse, France}}
\affiliation[40]{\small{ARTEMIS, UMR 7250 (CNRS/OCA/UNS), boulevard de l'Observatoire, BP 4229, F 06304 Nice Cedex, France}}
\affiliation[41]{\small{Observatoire de Haute-Provence, F-04870 Saint Michel L'Observatoire, France}}
\affiliation[42]{\small{University of Michigan, 500 East University, Ann Arbor, MI 48109-1120, USA}}
\affiliation[43]{\small{Department of Astronomy, University of California, Berkeley, CA 94720-3411, USA}}
\affiliation[44]{\small{Department of Physics and Astronomy, University of Leicester, Leicester, LE1 7RH, UK}}
\affiliation[45]{\small{NASA Goddard Space Flight Center, Greenbelt, MD 20771, USA}}
\affiliation[46]{\small{Department of Astronomy and Astrophysics, Pennsylvania State University, 525 Davey Lab, University Park, PA 16802, USA}}
\affiliation[47]{\small{School of Physics, University of Western Australia (UWA), Crawley WA 6009, Australia}}

\emailAdd{amathieu@cppm.in2p3.fr}
\emailAdd{dornic@cppm.in2p3.fr}

  \abstract{
High-energy neutrinos could be produced in the interaction of charged cosmic rays with matter or radiation surrounding astrophysical sources. Even with the recent detection of extraterrestrial high-energy neutrinos by the IceCube experiment, no astrophysical neutrino source has yet been discovered. Transient sources, such as gamma-ray bursts, core-collapse supernovae, or active galactic nuclei are promising candidates. 
Multi-messenger programs offer a unique opportunity to detect these transient sources. By combining the information provided by the ANTARES neutrino telescope with information coming from other observatories, the probability of detecting a source is enhanced, allowing the possibility of identifying a neutrino progenitor from a single detected event. 
A method based on optical and X-ray follow-ups of high-energy neutrino alerts has been developed within the ANTARES collaboration. This program, denoted as TAToO, triggers a network of robotic optical telescopes (TAROT and ROTSE) and the \textit{Swift}-XRT with a delay of only a few seconds after a neutrino detection, and is therefore well-suited to search for fast transient sources. To identify an optical or X-ray counterpart to a neutrino signal, the images provided by the follow-up observations are analysed with dedicated pipelines. A total of 42 alerts with optical and 7 alerts with X-ray images taken with a maximum delay of 24 hours after the neutrino trigger have been analysed. No optical or X-ray counterparts associated to the neutrino triggers have been found, and upper limits on transient source magnitudes have been derived. The probability to reject the gamma-ray burst origin hypothesis has been computed for each alert.}



\maketitle
\flushbottom


\section{Introduction}

High-energy neutrinos are expected to be produced in the interactions of accelerated charged cosmic rays with matter and radiation fields within and surrounding their astrophysical sources~\citep{Chiarusi,Anchordoqui}. Neutrinos are unique messengers for studying the high-energy Universe as they are neutral and stable, interact weakly, and travel directly from their source without absorption. Neutrinos easily escape the acceleration regions and propagate through space without interaction, unlike charged particles, which are deflected by magnetic fields and no longer point back to their source. Detection of high-energy neutrinos from an astrophysical source would be a direct signature for the presence of hadronic acceleration, and would therefore provide crucial information on the origin of very high-energy cosmic rays.

The production of high-energy neutrinos has been proposed to occur in several kinds of astrophysical sources, in which the acceleration of hadrons may occur, such as active galactic nuclei (AGN), gamma-ray bursts (GRBs), supernova remnants and microquasars.
Many of these astrophysical accelerators show transient behaviour. Variations in the energy output of the most powerful astrophysical objects cover a large range in the time domain, from seconds for GRBs~\citep{Kouveliotou} 
to weeks for AGN~\citep{FermiBlazars} or core-collapse supernovae (CCSNe)~\citep{CCSNe_variability}.
The particularity of these high-energy phenomena is that they radiate over the entire electromagnetic (EM) spectrum, from the radio domain to TeV gamma 
rays. Other approaches to astronomy do not use EM radiation at all however, but use other messengers, such as neutrinos, gravitational waves or cosmic rays. The detection of astrophysical sources of these other messengers is very difficult due to the very small number of expected events and a large background contamination. A way to overcome this difficulty is to combine the detection of non-EM messengers with the EM signal, to provide a multi-messenger dataset. 

The two largest operating large-volume neutrino telescopes in the world are: the Northern hemisphere ANTARES telescope~\citep{ANTARES}, located 
in the Mediterranean Sea, and the Southern hemisphere IceCube telescope~\citep{IceCube}, located in the South Pole ice. These telescopes are designed 
to search for high-energy cosmic neutrinos (E$_{\nu}>100$ GeV) generated in extreme astrophysical sources. Each detector is able to monitor a full hemisphere 
of the sky (or even the whole sky if downgoing events are considered) with a high duty cycle. The recent IceCube discovery of extraterrestrial high-energy neutrinos~\citep{IC_paper_1,IC_paper_2} opened new windows in the field of astroparticle physics. This result has fixed the scale of neutrino fluxes in the 
Universe and has inspired a large number of hypotheses for their origin, mainly due to the poor localisation of the events. Up to now, no high-energy neutrino source has been identified. 

Searches for transient astrophysical phenomena offer very promising opportunities for high-energy neutrino telescopes, because the relatively short duration of the events means that the level of contamination (from atmospheric muons and atmospheric neutrinos) is strongly reduced. Targeted searches performed by the ANTARES telescope on GRBs~\citep{GRB_ANTARES}, on AGN flares~\citep{AGN_ANTARES} and microquasar outbursts~\citep{microquasars_ANTARES} have so-far yielded limits on the neutrino production in these sources. Taking full advantage of the possibilities offered by multi-messenger searches for transient sources, a multi-wavelength follow-up program, denoted as TAToO (Telescopes-ANTARES Target of Opportunity), has operated within the ANTARES Collaboration since 2009~\citep{TAToO}. It is based on optical and, since mid 2013, X-ray follow-ups of selected high-energy neutrino events very shortly after their detection. The network is composed of the small robotic optical telescopes TAROT~\citep{TAROT_klotz} and ROTSE~\citep{ROTSE_akerlof}, and the \textit{Swift} X-ray telescope (XRT)~\citep{XRT}. 
This approach has the advantage that it does not require an \textit{a priori} hypothesis on the nature of the underlying neutrino source. To be sensitive to all types of time variability in the astrophysical sources, the observational strategy is composed of a real-time observation for rapidly fading sources, such as GRB afterglows, complemented by several observations during the following month specially adapted to detect the rising light curve of CCSNe. Unprecedented in this domain, the ANTARES telescope is able to generate alerts within a few seconds after the neutrino detection, and provides a precision on the reconstructed direction to better than 0.5$^{\circ}$ at high energies (E $>$ 1 TeV). The TAToO system is therefore well-suited to searching for rapid transients showing time variability at the minute scale. A similar program is running in IceCube \citep{IceCube_follow-up} since 2008, but with larger alert generatoin delay.

With this type of real-time analysis, one neutrino associated with a transient optical or X-ray counterpart would be a significant event. Since 2009, around 150 neutrino alerts have been sent to the optical telescope network. Among them, 42 alerts have had follow-up observations within one day after the neutrino detection. With the \textit{Swift}-XRT follow-up, a total of 7 alerts have been studied.

In this paper, the first results of the analysis of the rapid follow-up observations associated with the TAToO neutrino alerts are presented. A future dedicated paper will discuss the results of the analysis of the long-term follow-up. A brief description of the ANTARES experiment and the alert system is given in section 2. The neutrino data set used in this analysis is described in section 3. The optical and X-ray analyses, as well as the results are presented in sections 4 and 5 respectively. Finally, the implications of these results for GRB neutrino production models are discussed in section 6.


\section{TAToO and its multimessenger network}

	\subsection{The ANTARES neutrino telescope}
	
	For the ANTARES telescope~\citep{ANTARES}, events are detected underwater by the Cherenkov light induced in the medium by secondary particles resulting from neutrino interactions in the darkness of the deep sea. The detector is a 3D array of roughly 900 optical sensors in an approximately cylindrical volume of base diameter 190 m and height 350 m. Each optical sensor includes a 10-inch photomultiplier oriented downwards to maximise the detection efficiency to upgoing particles.

	Owing to their low interaction probability, neutrinos have the ability to cross the Earth, and muon neutrinos interacting close to the detector produce an upgoing muon. At high energy (E$_{\nu}\geq$ 1 TeV), the Lorentz boost is such that the muon and neutrino directions are identical to within a few tenths of a degree, so that accurately reconstructing the muon trajectory provides a good estimate of the arrival direction of the neutrino candidate. Another source of muons is due to cosmic rays that hit atmospheric nuclei and produce particle showers. These "atmospheric muons" have downgoing trajectories, which can therefore be used to discriminate against them. These muons, whose abundance at the ANTARES detector is roughly six orders of magnitude larger than those induced by atmospheric neutrinos, are the main background and have to be efficiently suppressed. The signal recorded by the detector is also polluted by decays of natural Potassium 40 in sea water and bioluminescence~\citep{ANTARES}. These backgrounds can be mitigated by applying causality criteria to the observed light patterns. 
	
	The telescope, fully equipped since May 2008, is anchored to the seabed at 2475 m below the Mediterranean sea surface, at 42$^\circ$48'N 6$^\circ$10'E, with a maximum efficiency in the Southern sky~\citep{point_source_ANTARES}. All signal hits from the photomultipliers are digitised in situ with a time resolution of about 1 ns, time stamped by a clock system synchronous to a standard GPS and sent to a 40 km distant shore station. When several optical sensors show a topology compatible with a muon track crossing the instrumented volume, all photomultiplier data is stored by the PC farm at the shore station and made available within a few seconds for further track reconstruction and physics analysis.


	\subsection{The TAToO alert system}
	\label{section:TAToO}
	
	The criteria for the TAToO trigger are based on the features expected from astrophysical sources. Several models predict the production of neutrinos with energy greater than 1 TeV from GRBs~\citep{GRBs1,GRBs2,GRBs3,GRBs4}, CCSNe~\citep{ccSNe} and AGN~\citep{AGN_hadron_model}. A basic requirement for the near coincident observation of a neutrino and an optical counterpart is that the pointing accuracy of the neutrino telescope should be at least comparable to the field of view of the TAROT and ROTSE telescopes ($\approx 2^\circ \times 2^\circ$). After the selection of upgoing events, which removes the huge background of atmospheric muons, the ANTARES neutrino sample consists mainly of atmospheric neutrinos. Different criteria are used to select candidates with an increased probability to be of cosmic origin ~\citep{TAToO}.
	A fast and robust algorithm is used to reconstruct the data~\citep{BBFit_ANTARES}, which uses an idealised detector geometry and is thus independent of the dynamical positioning calibration. This reconstruction, and a subsequent quality selection (Nline$\geq 2$, zenith angle $<$ 0, and $Q \leq 1.3+[0.04(N_{\text{hits}}-5)]^2$, see Ageron et al. 2012) allows the rate of events to be reduced from few Hz down to few mHz. The remaining events are then passed to a more precise reconstruction tool~\citep{AAFit_ANTARES} which allows the neutrino nature of the event to be confirmed, and the angular resolution to be improved. 
	
	Three online neutrino trigger criteria are currently implemented in the TAToO alert system:
	\begin{itemize}
		\item Doublet trigger: the detection of at least two neutrino-induced muons coming from similar directions ($< 3^\circ$) within a predefined time 
		window ($< 15$ minutes).
		\item High-energy trigger: the detection of a single high-energy ($\sim 7$ TeV) neutrino-induced muon.
		\item Directional trigger: the detection of a single neutrino-induced muon for which the direction points toward a local galaxy ($< 0.5^{\circ}$). The coordinates of the galaxies are selected from the GWGC catalogue~\citep{GWGC} with a distance cut at 20~Mpc.
	\end{itemize}
	
\begin{figure}[!h]
		\centering
		\includegraphics[trim={0 0 0 9cm},clip,width=0.65\textwidth]{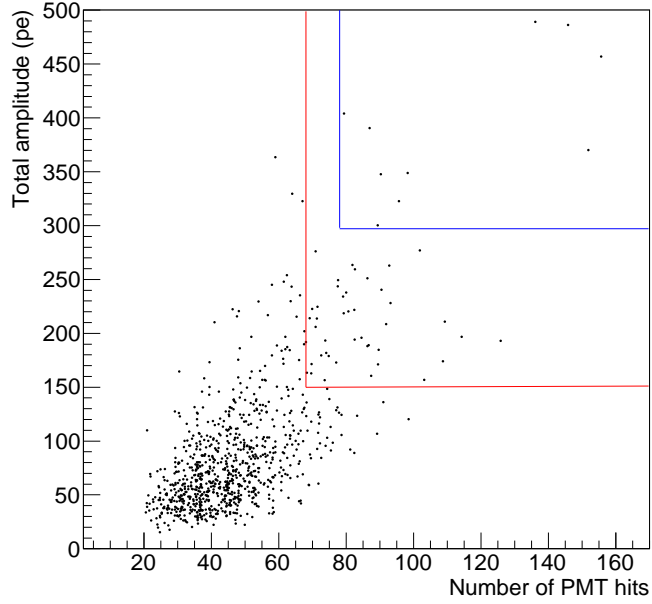}
		\caption{Scatter plot of the number of photomultiplier hits vs. the total amplitude (in photoelectrons) for neutrino candidates. The high-energy 
		trigger criterion is fulfilled by events inside the red box. Events inside the blue box correspond to the very high-energy trigger for the 
		\textit{Swift}-XRT.}
		\label{fig:amp_nhit_HE_alerts}
	\end{figure}

	In agreement with the optical telescopes, the total trigger rate is tuned to 25 per year. This rate is dominated by high-energy and directional triggers, as until now no doublet trigger has been sent to the network. The accidental coincidence rate due to two uncorrelated events is estimated to be $7 \times 10^{-3}$ per year. The high-energy trigger typically requires more than 70 photomulitplier hits and a total amplitude greater than 150 photoelectrons. To comply with the rate of 6 alerts per year required by the \textit{Swift} satellite, a subset of the high-energy trigger, denoted as the very high-energy trigger, provides a dedicated trigger for the XRT. It typically requires more than 80 hits and 300 photoelectrons. The bidimensional distribution of variables used for the selection of both high-energy and very high-energy triggers is illustrated in figure~\ref{fig:amp_nhit_HE_alerts} for neutrino candidates recorded from 2012 to March 2015. Neutrinos falling outside the red box can lead to an alert if they fulfil the directional or the doublet trigger. The directional trigger was implemented in late 2011. The performances of these three triggers are described in table~\ref{tab:perf_tatoo}.

	\begin{table}
	\centering
	\caption{Performances of the three alert criteria.}           
	\label{tab:perf_tatoo}   
	\begin{threeparttable}[b]
	\begin{footnotesize}
	\begin{tabular}{c c c c c}
	\hline\hline
	\noalign{\smallskip}
	Trigger 	 & Angular Resolution (median)   & PSF coverage\tnote{a}     & Atmospheric muon contamination 	& Mean energy\tnote{b}  	 \\  
	\noalign{\smallskip}    
	\hline
	\noalign{\smallskip}  	
	High energy        &  $0.25-0.3^{\circ}$              &     96~\%  &  $< 0.1~\% $           &  $\sim 7$ TeV  \\
	Directional   &  $0.3-0.4^{\circ}$           &     90~\%  &  $\sim 2$~\%                &  $\sim 1$ TeV  \\
	Doublet          &           $\leq~0.7^{\circ}$           &                               &    0~\%               & $\sim 100$ GeV  \\
	\noalign{\smallskip}
	\hline
	\end{tabular}
	\begin{tablenotes}
	\item[a] Percentage of the ANTARES PSF covered by a $2^\circ \times 2^\circ$ telescope field of view, assuming events produced in GRBs.
	\item[b] Neutrino energy weighted assuming the atmospheric muon neutrino spectrum.
	\end{tablenotes}
	\end{footnotesize}
	\end{threeparttable}
	\end{table}
	
\begin{figure}[!h]
	\centering
	\includegraphics[scale=0.4,trim={0.8cm 0 0 0},clip]{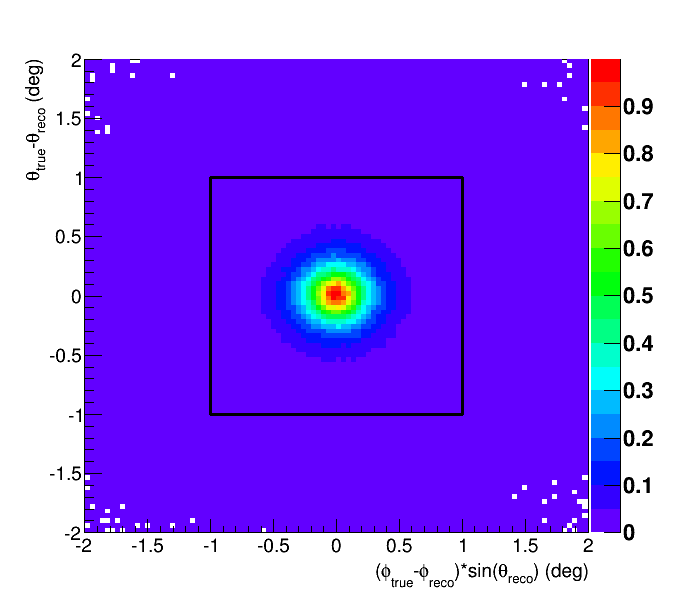}
	\caption{Bi-dimensional angular resolution for a typical high-energy neutrino alert. The black square corresponds to the TAROT and ROTSE telescope 
	field of view ($\approx 2^\circ \times 2^\circ$).}
	\label{fig:PSF_TAToO}
	\end{figure}

	Figure~\ref{fig:PSF_TAToO} shows the estimate of the point spread function (PSF) for a typical high-energy neutrino alert, compared to the field of view of 
	TAROT and ROTSE. For the highest energy events, the angular resolution reaches less than 0.3$^\circ$ (median value). 

\begin{figure}[!h]
		\centering
		\includegraphics[width=0.68\textwidth]{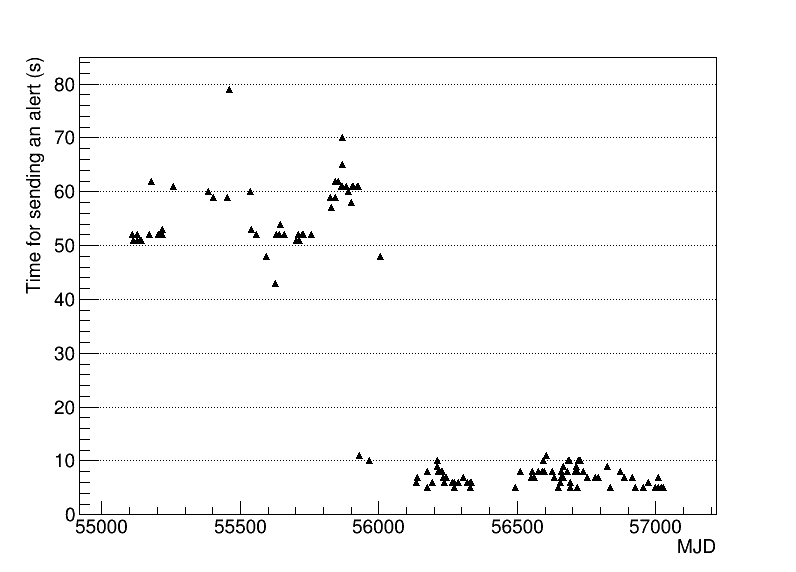}
		\caption{Latency for the 150 triggers sent by the alert system. The step from 50 to 5 seconds 
		corresponds to an upgrade of the ANTARES DAQ system.}
		\label{fig:AlertLatency}
	\end{figure}

	Before 2012, the alert system sent alerts about 50 seconds after the neutrino events. After 2012, a major DAQ system improvement enabled ANTARES to send alerts a few seconds ($\sim 3-5$ seconds) after the detection of the neutrinos. Figure~\ref{fig:AlertLatency} displays this latency for the 150 alerts collected since the commisioning of the TAToO alert system.
			
\section{Neutrino data set}

	This dataset is composed of 42 alerts with an early optical follow-up and 7 alerts with an X-ray follow-up. Only one alert has an early follow-up with both optical and X-ray telescopes. The direction, time and energy of the alerts are reported in table~\ref{tab:neutrino_data}. A total of 34 and 14 neutrinos have been triggered as high-energy and directional alerts respectively. These numbers are in agreement with expectations from Monte Carlo simulation assuming that they are of atmospheric origin.

\begin{figure}[!h]
	\centering
	\includegraphics[trim={0 0 0 18cm},clip,width=0.78\textwidth]{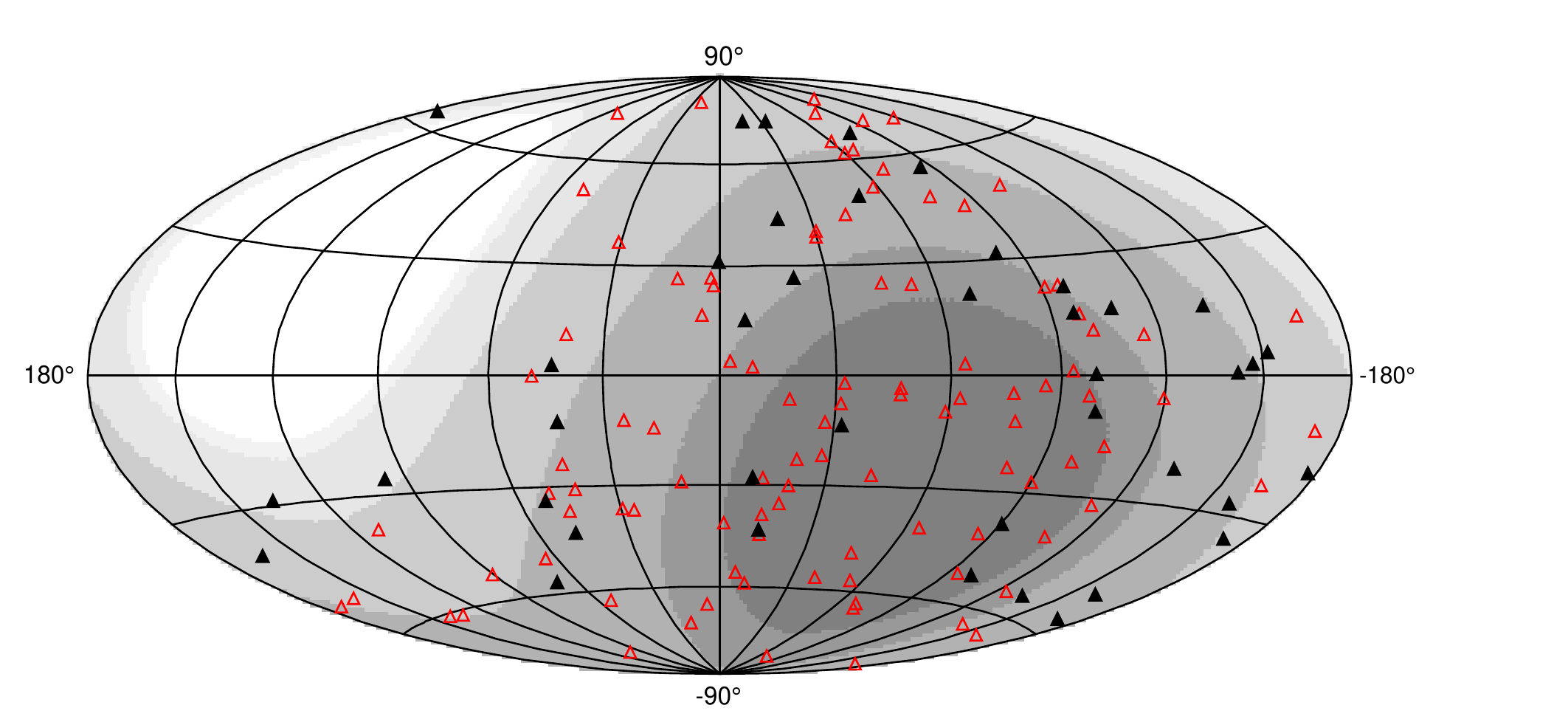}
	\caption{Skymap in galactic coordinates of the candidates which triggered the alert system. Black triangles refer to the triggers which had an early optical follow-up, while red triangles show the triggers with no early observation. The visibility of the ANTARES detector is represented in greyscale, with a visibility of 100~\% for the darkest zone and a white region with no visibility.}
	\label{fig:SkyMapGal}
	\end{figure}

	A skymap in galactic coordinates of the neutrino alert directions which have triggered the alert system since 2009 is displayed in figure~\ref{fig:SkyMapGal}.

	\begin{table}
	\centering
	\caption{Details of the 48 neutrinos which triggered the TAToO alert system and for which early optical and/or X-ray images were taken.}           
	\label{tab:neutrino_data}  
	\begin{threeparttable}[b]  
	\begin{footnotesize}
	\begin{tabular}{c c c c c c c c}
	\hline\hline
	\noalign{\smallskip}
	Alert name 	 &   Trigger time  &  Ra   &   Dec   &   Trigger type\tnote{a}   &   Nhits   &   Amp\tnote{b}   &   Nline\tnote{c}  	 \\ 
	(ANTyymmddA/B) & (UT) & ($^\circ$) & ($^\circ$) & & & & \\ 
	\noalign{\smallskip}    
	\hline
	\noalign{\smallskip}  	
	ANT100123A & 07:16:45 & 73.071 & -2.991 & HE & 34 & 339 & 4 \\
	ANT100725A & 01:16:40 & 257.224 & -63.188 & HE & 12 & 66 & 3 \\
	ANT100913A & 21:55:25 & 165.142 & 41.391 & HE & 46 & 277 & 5 \\
	ANT100922A & 11:24:23 & 43.578 & 13.458 & HE & 19 & 158 & 3 \\
	ANT110305A & 12:57:20 & 295.517 & -48.193 & HE & 27 & 194 & 8 \\
	ANT110409A & 04:47:50 & 129.365 & -40.207 & HE & 21 & 253 & 8 \\
	ANT110531A & 21:45:16 & 326.944 & -8.795 & HE & 54 & 439 & 8 \\
	ANT110923A & 08:26:33 & 285.194 & 10.244 & HE & 22 & 236 & 7 \\
	ANT110925B & 20:27:38 & 70.998 & 15.507 & HE & 50 & 356 & 11 \\
	ANT111008A & 10:45:40 & 248.170 & -25.834 & HE & 36 & 336 & 9 \\
	ANT111019A & 08:56:15 & 321.165 & -0.692 & HE & 32 & 185 & 8 \\
	ANT111019B & 08:56:15 & 322.070 & -0.637 & HE & 32 & 185 & 8 \\
	ANT111101A & 12:30:52 & 345.917 & -5.698 & HE & 33 & 207 & 6 \\
	ANT111205A & 21:33:11 & 146.411 & -32.829 & Dir & 20 & 74 & 9 \\
	ANT111228A & 09:18:42 & 27.025 & 32.323 & Dir & 11 & 90 & 7 \\
	ANT120102A & 01:24:21 & 72.129 & -59.709 & Dir & 11 & 54 & 8 \\
	ANT120105A & 18:02:53 & 228.824 & -26.121 & HE & 26 & 184 & 7 \\
	ANT120730A & 00:14::33 & 178.088 & -40.096 & HE & 135 & 489 & 11 \\
	ANT120907A & 01:25:10 & 220.171 & -11.595 & HE & 77 & 245 & 10 \\
	ANT120907B & 10:12:41 & 344.750 & 30.933 & HE & 76 & 242 & 9 \\
	ANT120923A & 10:00:16 & 318.596 & -51.071 & HE & 71 & 128 & 4 \\
	ANT121010A & 06:31:01 & 52.841 & -29.216 & Dir & 38 & 28 & 7 \\
	ANT121012A & 04:42:49 & 239.512 & -10.226 & Dir & 47 & 42 & 9 \\
	ANT121027A & 18:47:58 & 105.552 & -4.011 & Dir & 39 & 51 & 7 \\
	ANT121206A & 10:02:42 & 62.001 & 2.434 & HE & 104 & 158 & 10 \\
	ANT130210A & 14:00:10 & 185.433 & 5.897 & Dir & 52 & 61 & 8 \\
	ANT130722A & 20:38:24 & 74.570 & 3.411 & VHE & 87 & 188 & 9 \\
	ANT130724A & 02:07:20 & 199.295 & 11.938 & HE & 91 & 171 & 10 \\
	ANT130915A & 06:24:58 & 311.314 & -51.603 & VHE & 146 & 485 & 10 \\
	ANT130927A & 18:22:56 & 106.948 & -68.578 & VHE & 81 & 258 & 10 \\
	ANT130928A & 20:03:18 & 121.865 & -3.156 & Dir & 31 & 104 & 4 \\
	ANT131027A & 12:13:35 & 105.124 & 6.422 & HE & 74 & 190 & 7 \\
	ANT131209A & 01:56:24 & 197.235 & -13.345 & HE & 87 & 163 & 9 \\
	ANT131221A & 08:43:36 & 138.015 & -23.930 & Dir & 30 & 49 & 9 \\
	ANT140123A & 11:53:45 & 105.440 & 0.811 & VHE\tnote{*} & 80 & 401 & 8 \\
	ANT140125A & 03:27:08 & 53.002 & -49.491 & HE & 80 & 224 & 8 \\
	ANT140203A & 06:48:06 & 89.731 & -23.394 & Dir & 54 & 93 & 9 \\
	ANT140223A & 01:01:20 & 43.712 & -10.802 & Dir & 39 & 30 & 6 \\
	ANT140304A & 05:19:15 & 117.930 & -44.944 & HE & 81 & 192 & 9 \\
	ANT140309A & 00:40:10 & 52.511 & -12.955 & HE & 71 & 275 & 7 \\
	ANT140311A & 13:16:20 & 237.416 & 31.826 & VHE & 96 & 324 & 11 \\
	ANT140323A & 15:31:01 & 150.919 & -27.371 & Dir & 46 & 102 & 6 \\
	ANT140408A & 11:10:09 & 177.461 & -3.301 & Dir & 44 & 70 & 7 \\
	ANT140505A & 11:54:47 & 203.359 & 14.450 & HE & 80 & 239 & 9 \\
	ANT140914A & 05:13:35 & 298.424 & 2.677 & HE & 109 & 210 & 9 \\
	ANT141220A & 14:53:59 & 71.679 & -45.968 & VHE & 59 & 361 & 6 \\
	ANT150122A & 07:09:35 & 168.551 & -26.119 & Dir & 28 & 55 & 7 \\
	ANT150129A & 10:20:12 & 34.656 & 4.863 & VHE & 92 & 207 & 10 \\
	\noalign{\smallskip}
	\hline
	\end{tabular}
	\begin{tablenotes}
	\item[a]HE: high energy, Dir: directional, VHE: very high energy. With this last trigger, alerts are sent to the optical telescope network and to the \textit{Swift}-XRT.
	\item[b] Amplitude in photoelectrons.
	\item[c] Number of ANTARES lines on which the neutrino event has been detected.
	\item[*] Early observations by both the optical telescopes and the \textit{Swift}-XRT have only been made for this trigger.
	\end{tablenotes}
	\end{footnotesize}
	\end{threeparttable}
	\end{table}


\section{Optical follow-up}

	\subsection{The optical telescope network}

	The optical follow-up of ANTARES alerts is performed using ground-based telescopes dedicated to early observations of GRBs. Six such telescopes are involved: ROTSE-III (USA, Namibia, Australia, Turkey) and TAROT (France and Chile). ROTSE-III is a network of four identical 0.45 m telescopes~\citep{ROTSE_akerlof} and TAROT is a network of two identical 0.25 m telescopes~\citep{TAROT_klotz}. ROTSE has stopped its activity progressively over the last years: ROTSE 3a (Australia), 3b (Texas), 3c (Namibia) and 3d (Turkey) have been stopped in 07/2011, 09/2014, 12/2010 and 12/2012, respectively. These telescopes had a field of view of $1.9^{\circ} \times 1.9^{\circ}$ and a spatial sampling of 3.3 arcsec/pixel. The sensitivity of the TAROT and ROTSE telescopes is about the same: for a signal to noise ratio of 5, an exposure time of 180 seconds and a clear filter, the limiting magnitude is $\sim 18.5$. 
	
\begin{figure}[!h]
		\centering
		\includegraphics[scale=0.55,trim={0 0 3cm 18cm},clip]{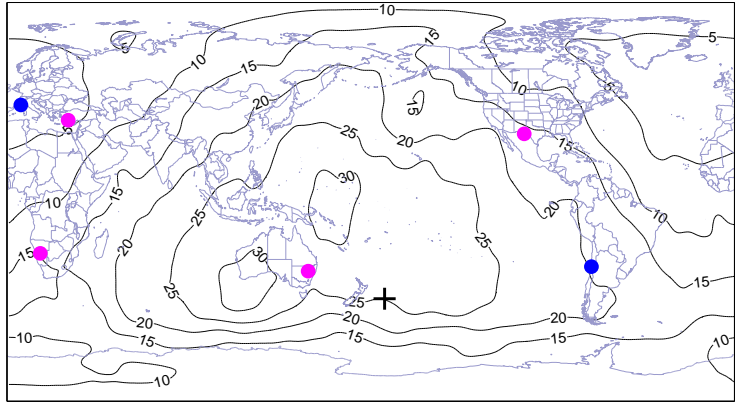}
		\caption{The contours of the world map indicate the percentage of neutrino triggers visible immediately (based on 140 ANTARES alerts). Blue and magenta points are the locations of TAROT and ROTSE telescopes respectively. The black cross indicates the antipodal point of the ANTARES experiment.}
		\label{fig:map_tarot_rotse}
	\end{figure}

	Alerts are sent automatically by an ANTARES server using a socket protocol, similar to the Gamma-ray bursts Coordinates Network (GCN) system developed by~\cite{GCN_barthelmy} for GRBs. If we consider that ANTARES events originate from GRBs, the rapid decay of their afterglow light implies the use of telescopes with very fast slewing. The time reaction of these optical telescopes is of the order of a few seconds when the field of view is observable. Figure~\ref{fig:map_tarot_rotse} shows a map of the real-time visibility of ANTARES neutrino alerts by a single telescope as a function of its location on Earth.
	
For the prompt observations, the efficiency of an optical ground-based telescope depends on the object visibility (night and field elevation) and the local conditions (weather and technical problems). For instance, among 744 GRBs triggered by \textit{Swift} between September 2006 and January 2015, only 18~\% (i.e. 134) were accessible during the prompt emission at TAROT Chile. Only 45 GRBs were really observed during this period. This implies an efficiency of about 33~\% for TAROT Chile site (i.e. weather and technical problems). As a comparison, for the ANTARES alerts, figure~\ref{fig:map_tarot_rotse} shows that only 20~\% of triggers are promptly accessible from TAROT Chile (i.e. 30 among 150). Table~\ref{tab:table_optical_alerts} shows that only 11 ANTARES triggers were really observed in less than a few minutes by TAROT Chile, indicating a similar efficiency for ANTARES and \textit{Swift} triggers. The combination of visibility, weather, and reliability of TAROT Chile allows the prompt observation of one trigger every 15. This emphasizes the necessity of using a telescope network to cover a prompt follow-up of the ANTARES triggers.

	The early observation strategy of TAROT consists of taking 6 images of 180 seconds exposure as fast as possible while for ROTSE, up to 30 images of 20 to 60 seconds are taken (depending on the presence of the Moon in the observed sky). Each image is preprocessed directly on site and then transferred to the ANTARES data storage for the image analysis.

	\subsection{Optical image analysis}

\begin{figure}[!h]
		\centering
		\includegraphics[width=0.75\textwidth]{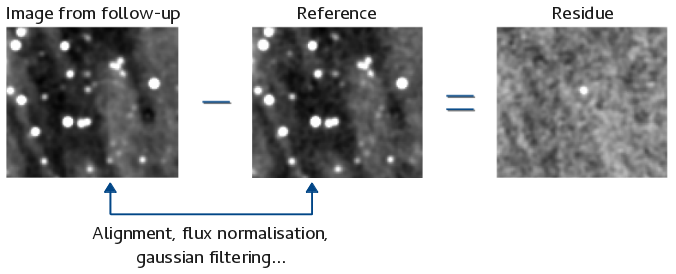}
		\caption{Optical image analysis based on the subtraction method. A good reference image \textit{(center)} 
		is subtracted from an image \textit{(left)} where the counterpart signal is expected. Positive residues 
		in the subtracted image \textit{(right)} are then analysed.}
		\label{fig:sub_method}
	\end{figure}

	To search for a transient optical counterpart to the neutrino trigger, images provided by the follow-up of neutrino alerts must be processed and analysed (TAROT and ROTSE images are always analysed separately). For this, a new pipeline based on an image subtraction method has been developed (figure~\ref{fig:sub_method}). The first step of this program is the astrometric and photometric calibration of images.

	A necessary step of the image subtraction method is to calibrate and align all the images. To calibrate astrometry, \textit{SExtractor} \citep{SExtractor} is used to extract sources from an image and to build a catalogue containing the coordinates of each detected source. The extraction threshold is set at 2.5~$\sigma$ above background noise. This catalogue and the information provided by the image header (frame dimensions, World Coordinate System data, etc.) are then read by \textit{SCAMP}~\citep{SCAMP} and compared with a reference catalogue to align the images.

	The photometric calibration is done with \textit{LePHARE} \citep{LePHARE}. This program can compute theoretical magnitudes for different bands by using a reference photometric catalogue and a list of filter response curves. The photometric catalogue is provided by NOMAD~\citep{NOMAD} and contains magnitudes in B, V, R, J, H, K bands for the brightest stars in the field of the image. Knowing the charge-coupled device (CCD) response curve, magnitudes in the CCD band\footnote{All our images are taken with the clear filter.} can be computed. 

	After calibration, one can select a reference image (REF) to perform the subtraction. This reference must be a very good quality image (in terms of limiting magnitude) with no or less signal than in the image to analyse. As we are looking for transient sources with a high variability on the minute/hour timescale, only images (IM) taken less than 24 hours after the neutrino trigger are analysed. 
	Therefore, a reference can be chosen among images taken several days or weeks after the alert. Some corrections are then applied to images (e.g. gaussian filter, flux normalisation) and subtraction IM-REF is made pixel by pixel using the \textit{imarith} function from the \textit{IRAF} package \citep{IRAF}. 
	Counterpart candidates are automatically selected among positive residual objects on the subtracted image if they satisfy one of the two following 
	criteria: 
	\begin{itemize}
		\item New source: the candidate is detected in IM and not in REF.
		\item Magnitude variation: the source in IM is at least 0.5 magnitude brighter than in REF.
	\end{itemize}

	All these candidates are then checked by eye and light curves are built for the interesting ones. Only candidates with at least two points in the light curve (i.e. candidates detected in at least two images), or one very bright point (i.e. $\sim 2$ mag below the limiting magnitude) are considered. Asteroids and variable stars are rejected.
	If a candidate showing variability (for example, a fast decrease as typical for GRB afterglows) is not already catalogued, it is considered as a strong 
	candidate to be the counterpart.

\begin{figure}
	 	\centering
	 	\includegraphics[scale=0.45]{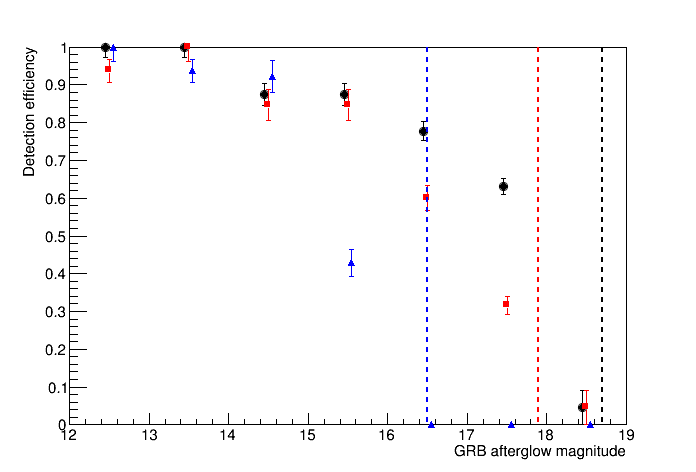}
	 	\caption{Detection efficiency of the optical image analysis pipeline estimated by inserting fake GRB afterglows 
	 	into three sets of images with limiting magnitudes (vertical dashed lines) of 16.5 (blue), 17.9 (red) and 18.7 
	 	(black) and considering a decay of the light curve $\propto t^{-1}$.}
	 	\label{fig:detection_efficiency_GRB}
	 \end{figure}

	 The detection efficiency of the image analysis pipeline has been estimated by injecting fake GRB afterglows in three different observations consisting of 6 images of 180 seconds exposure each. The mean limiting magnitude of these images is 16.5, 17.9 and 18.7 for the three observations, respectively. 
	 Assuming a light curve decay $\propto t^{-1}$, 125 GRB afterglows between magnitude 12 and 19 have been inserted in each of the three sets of images. 
	 The efficiency, presented in figure~\ref{fig:detection_efficiency_GRB}, is given by the fraction of inserted GRB afterglows that has been detected by 
	 the processing and the candidate identification.
	 The efficiency of the pipeline is better than 95~\% for bright GRB afterglows (magnitude between 12 and 14) and reaches $\sim 50$~\% around 1 magnitude below 
	 the limiting magnitude of images (vertical dashed lines).
	 
	\subsection{Optical follow-up results}
	    
    A total of 42 alerts from January 2010 to January 2015 with early images have been analysed. The telescopes that observed the regions around the neutrino direction, the number of early images analysed, their exposure as well as the delay and the limiting magnitude of the first image, are given in 
    Table~\ref{tab:table_optical_alerts}. 
    For 11 alerts, the time between the neutrino detection and the start of the acquisition of the first image is less than 1.2 minutes and is as low as 25 
    seconds for 7 alerts. 
    
    No optical counterpart associated with one of the 42 neutrinos was found. This could indicate that the observations were made too late when the sources could 
    have faded below the sensitivity of the optical telescopes or that the sources occurred outside the field of view of the telescopes, or that the neutrinos that generated the triggers have another origin.
    Upper limits on the magnitude of possible transient sources which could have emitted the neutrino have been derived. These limits 
    correspond to the limiting magnitude of images that is the faintest signal that can be detected. 
    As we are looking for rapidly-fading sources, the signal is supposed to be stronger in the first image of the observation, so the upper limits 
    are the limiting magnitude of each first image computed at the 5~$\sigma$ level and corrected for galactic extinction~\citep{Schlegel} 
    (table~\ref{tab:table_optical_alerts}).
    
	\begin{table} 
		\centering 
		\caption{Details of the 42 neutrino alerts for which early optical images have been taken.}
		\label{tab:table_optical_alerts}  
		\begin{threeparttable}[b]
		\begin{footnotesize}
		\begin{tabular}{ccccccccc}
		\hline
		\hline
		\noalign{\smallskip}
		Alert name & Telescope & Analysed & Exposure \tnote{a} & Delay \tnote{b} & M$_{\text{lim}}$ \tnote{c} & $A_v$ \tnote{d} & Reference\tnote{e} & $P_{\text{reject}}^{GRB,\nu}$ \tnote{f} \\
		(ANTyymmddA/B)    &            &  images  & (sec)    &                                           &   (mag)   & (mag) &  (yymmdd) & \\
		\noalign{\smallskip}
		\hline
		\noalign{\smallskip}
		ANT100123A & TAROT & 6  & 180 & 17h47m  & 15.3 & 0.2 & 100211 & 0.00    \\ 
		ANT100725A & TAROT & 6  & 180 & 1m17s   & 16.1 & 0.3 & 100727 & 0.50 \\ 
				   & ROTSE & 30 & 20  & 1m15s   & 13.1 & 0.3 & 100731 & 0.12 \\ 
		ANT100913A & TAROT & 6  & 180 & 11h24m  & 17.6 & 0.0 & 100914 & 0.06 \\ 
		ANT100922A & ROTSE & 26 & 20  & 1h08m   & 13.6 & 0.5 & 101202 & 0.00    \\ 
		ANT110305A & ROTSE & 29 & 60  & 4h19m   & 15.7 & 0.1 & 110315 & 0.06 \\ 
		ANT110409A & TAROT & 6  & 180 & 1m08s   & 12.6 & 5.6 & 110423 & 0.04 \\ 
		ANT110531A & TAROT & 6  & 180 & 12h34m  & 17.6 & 0.1 & 110628 & 0.06 \\ 
		ANT110923A & TAROT & 7  & 180 & 9h58m   & 12.8 & 3.9 & 111001 & 0.00    \\ 
		ANT110925B & TAROT & 6  & 180 & 2h01m   & 15.2 & 1.8 & 111125 & 0.10 \\ 
				   & ROTSE & 30 & 60  & 50m58s  & 13.9 & 1.8 & 110927 & 0.00   \\ 
		ANT111008A & TAROT & 5  & 180 & 12h53m  & 14.3 & 2.5 & 111009 & 0.00    \\ 
		ANT111019A & ROTSE & 8  & 60  & 18h22m  & 16.7 & 0.1 & 111029 & 0.02 \\ 
		ANT111019B & ROTSE & 8  & 60  & 19h09m  & 16.9 & 0.1 & 111024 & 0.02 \\ 
		ANT111101A & ROTSE & 8  & 60  & 13h33m  & 17.2 & 0.1 & 111130 & 0.02 \\ 
		ANT111205A & TAROT & 6 & 180 & 10h05m  & 18.2 & 0.4 & 111207  & 0.16 \\ 
		ANT111228A & TAROT & 6 & 180 & 7h44m   & 17.0 & 0.1 & 120124  & 0.04 \\ 
				   & ROTSE & 8  & 60  & 7h53m   & 16.6 & 0.1 & 111231 & 0.04 \\ 
		ANT120102A & TAROT & 4 & 180 & 1m17s   & 17.0 & 0.1 & 120129  & 0.60 \\ 
		ANT120105A & ROTSE & 8  & 60  & 17h39m  & 16.0 & 0.4 & 120112 & 0.02 \\ 
		ANT120730A & TAROT & 26 & 180 & 20s     & 16.9 & 0.4 & 120807 & 0.88 \\ 
		ANT120907A & TAROT & 14 & 180 & 9m53s   & 15.9 & 0.2 & 120909 & 0.31 \\ 
		ANT120907B & TAROT & 11 & 180 & 18h15m  & 17.2 & 0.2 & 121106 & 0.02 \\ 
				   & ROTSE & 27 & 60  & 8h28m   & 15.9 & 0.2 & 121004 & 0.02 \\ 
		ANT120923A & TAROT & 6  & 180 & 15h43m  & 18.0 & 0.1 & 121020 & 0.03 \\ 
		ANT121010A & TAROT & 24 & 180 &  25s   & 18.6 & 0.0 & 121209  & 0.90 \\ 
		ANT121012A & TAROT & 6 & 180 &  19h06m & 16.5 & 0.7 & 121016  & 0.02 \\ 
		ANT121027A & ROTSE & 8 & 20  & 14h56m  & 13.4 & 2.6 & 121124  & 0.00    \\ 
		ANT121206A & ROTSE & 27 & 60  & 27s     & 15.6 & 1.1 & 121210 & 0.62 \\ 
		ANT130210A & ROTSE & 8 & 60  & 14h46m  & 16.5 & 0.1 & 130216  & 0.02 \\ 
		ANT130724A & TAROT & 3  & 180 & 18h04m  & 15.9 & 0.1 & 130729 & 0.02 \\ 
		ANT130928A & ROTSE & 8 & 60  & 13h49m  & 15.9 & 0.1 & 131006  & 0.02 \\ 
		ANT131027A & ROTSE & 8  & 20  & 18h14m  & 15.0 & 0.7 & 131112 & 0.00    \\ 
		ANT131209A & TAROT & 6  & 180 & 1h14m   & 16.3 & 0.1 & 131215 & 0.14 \\ 
		ANT131221A & TAROT & 2 & 180 & 18s     & 16.8 & 0.5 & 131222  & 0.83 \\ 
		ANT140123A & TAROT & 23 & 180 & 13h21m  & 16.2 & 1.3 & 140126 & 0.02 \\ 
		ANT140125A & TAROT & 6  & 180 & 1h14m   & 18.1 & 0.0 & 140128 & 0.43 \\ 
		ANT140203A & ROTSE & 8 & 60  & 19h43m  & 14.9 & 0.1 & 140205  & 0.00    \\ 
		ANT140223A & TAROT & 3 & 180 & 17h08m  & 15.9 & 0.1 & 140224  & 0.02 \\ 
				   & ROTSE & 3  & 60  & 31m29s  & 14.2 & 0.1 & 140227 & 0.02 \\ 
		ANT140304A & TAROT & 18 & 180 & 25s     & 17.5 & 0.6 & 140304 & 0.92 \\ 
		ANT140309A & TAROT & 16 & 180 & 24s     & 17.1 & 0.1 & 140309 & 0.88 \\ 
		ANT140323A & ROTSE & 8 & 60  & 14h47m  & 16.2 & 0.2 & 140325  & 0.02 \\ 
		ANT140408A & TAROT & 6 & 180 & 16h11m  & 16.7 & 0.1 & 140410  & 0.02 \\ 
				   & ROTSE & 8  & 60  & 19h07m  & 16.0 & 0.1 & 140410 & 0.02 \\ 
		ANT140505A & ROTSE & 2  & 60  & 17h11m  & 14.7 & 0.1 & 140510 & 0.00    \\ 
		ANT140914A & TAROT & 13 & 180 &  1m05s  & 16.9 & 0.5 & 140915 & 0.62 \\ 
		ANT150122A & TAROT & 8 & 180 &  17s     & 18.6 & 0.1 & 150122 & 0.90 \\ 
		\noalign{\smallskip}
		\hline
		\end{tabular}
		\begin{tablenotes}
		\item[a] Exposure of each image.
		\item[b] Delay in hours, minutes and/or seconds between the neutrino trigger and the first image.
		\item[c] Limiting magnitude of the first image computed at the 5~$\sigma$ level and corrected for the galactic extinction.
		\item[d] Galactic extinction from \cite{Schlegel}.
		\item[e] Date of the reference image used for the subtraction.
		\item[f] Probability to reject the GRB origin hypothesis for this neutrino trigger (see section~\ref{sec:optical_grb_assoc}).
		\end{tablenotes}
		\end{footnotesize}
		\end{threeparttable}
	\end{table}


\section{X-ray follow-up}

	\subsection{The \textit{Swift} X-ray telescope}
		
\begin{figure}[!h]
		\centering
		\includegraphics[scale=0.4,trim={0.8cm 0 0 0},clip]{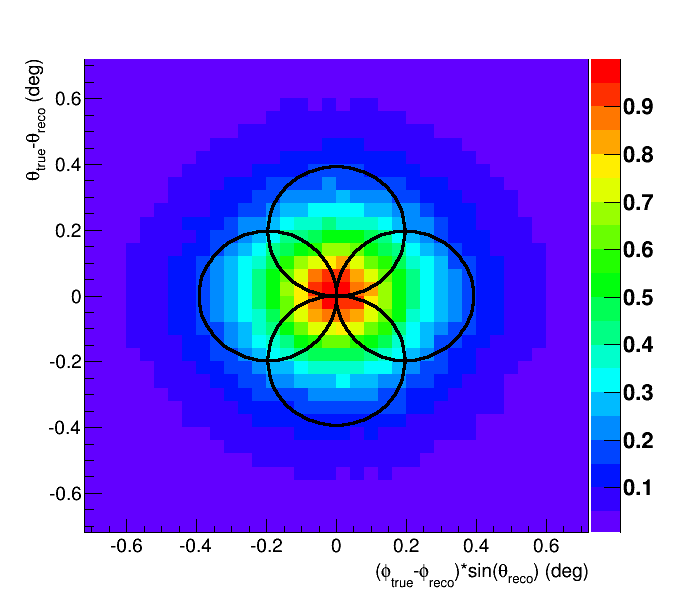}
		\caption{Each \textit{Swift}-XRT observation of an ANTARES trigger consists of 4 tiles (black circles), which covers an area of radius of $\sim 0.4^\circ$. With such a mapping, 72~\% of the the bi-directional uncertainty of a TAToO alert is covered.}
		\label{fig:4tiles_PSF_TAToO}
	\end{figure}

	The \textit{Swift} satellite \citep{Swift} with its XRT~\citep{XRT} provides a unique opportunity to observe X-ray counterparts to neutrino triggers. The detection sensitivity of the XRT is $5 \times 10^{-13}$ erg cm$^{-2}$ s$^{-1}$ in 1 ks, with an energy band covering from 0.3 to 10 keV. The follow-up of ANTARES alerts with the \textit{Swift}-XRT~(see section~\ref{section:TAToO}) was implemented in June 2013. Due to the small field of view (radius $\sim 0.2^\circ$) of the XRT and the typical error radius of an ANTARES alert ($\sim 0.3 - 0.4^\circ$), each observation is composed of 4 tiles up to 2 ks exposure each. This mapping covers about 72~\% of the ANTARES PSF for a very high-energy neutrino (figure~\ref{fig:4tiles_PSF_TAToO}). The choice of these 4 tiles corresponds to a trade-off between the requested exposure time and the neutrino PSF coverage. With an exposure of 2 ks, a sensitivity of $2.5 \times 10^{-13}$ erg cm$^{-2}$ s$^{-1}$ can be reached~\citep[][figure 14]{Evans_2014}. For each trigger, a single Automatic Target is uploaded, and XRT observations start as soon as possible. \textit{Swift} automatically divides the visibility window of the ANTARES trigger location in each spacecraft orbit between the 4 tiles. This is repeated on each orbit until the requested exposure time has been gathered for each tile. An automatic analysis of the data, described in the following section, is performed soon after, and in the case where an interesting candidate counterpart is found, further observations are scheduled.
	
	\subsection{X-ray data analysis}

	Each 4-tile observation with the \textit{Swift}-XRT is automatically analysed at the UK Swift Science Data Centre. The first step of the analysis is to combine the 4 tiles into a single image and to apply the source detection algorithm which has been developed for the 1SXPS catalogue~\citep{Evans_2014}. It consists of filtering the data, creating exposure maps, locating and characterising sources. The source positions are generally determined using the on-board star trackers with an accuracy of 3.5 arcsec \citep{Moretti}. However it is sometimes possible to enhance the astrometry either by using stars in the field of view of the UV/optical telescope~\citep{Goad,Evans_2009} or by matching the detected XRT sources with the 2MASS catalogue~\citep{Evans_2014}. The position with the smallest error is reported.\\
	A detection flag is then assigned to each detected source to determine the probability of the source being spurious. This flag can be either \textit{Good} (0.3~\% of sources flagged as \textit{Good} are spurious), \textit{Reasonable} (7~\% of false positive) or \textit{Poor} (35~\% of false positive). More details can be found in \cite{Evans_2014}.		

	Light curves and spectra are also built for each detected source using the tools described by \cite{Evans_2007,Evans_2009}. A cross-correlation with the X-ray master catalogue, the 1SXPS catalogue and SIMBAD\footnote{www.simbad.u-strasbg.fr} is then performed. The results of this processing, giving details of the newly discovered X-ray source(s) and any catalogue matches, are presented in webpages dedicated to each ANTARES alert.

	Finally, two tests are automatically carried out on each uncatalogued X-ray source to determine if it could be the counterpart to the ANTARES trigger.
	The first test relies on brightness and variability measurements: if a source is so bright that it should have been discovered previously\footnote{The mean XRT count-rate must be more than 1~$\sigma$ above the Rosat All-Sky Survey (RASS) 3~$\sigma$ upper limit~\citep{RASS} calculated at the location of the X-ray source. The Rosat upper limit is derived using RASS images, exposure maps and background maps.}, it means that the source is assumed to be new and therefore potentially the counterpart. The other possibility is if the source shows variability, and in particular significant signs of fading\footnote{The last bin of the source light curve is at least 3~$\sigma$ below the first bin.}, it is considered as a potential counterpart. The second test is based on the probability that the uncatalogued X-ray source is unrelated to the neutrino trigger. The number of expected serendipitous sources with a count-rate at least that of the detected source in each 4-tile observation is calculated by combining the minimum exposure necessary to detect the XRT source, the sky area, the expected sky density of sources and the completeness of the detection system. Details can be found in \cite{Evans_IceCube_2015}.


    \subsection{Results}
    
	The \textit{Swift}-XRT has responded to 7 neutrino triggers from the ANTARES detector between mid 2013 and the beginning of 2015 (table~\ref{tab:table_alerts_swift}). These 7 alerts have been observed after a mean delay of 3.6 hours in 4 tiles with exposure from 0.8 to 1.9 ks. With such exposure times, a sensitivity from 6.0 to 2.6$\times 10^{-13}$ erg cm$^{-2}$ s$^{-1}$ can be reached, respectively.
	
	The tiled analysis found 20 X-ray sources, among which 2 sources were already known (see table~\ref{tab:table_sources_swift}). Although 18 new X-ray sources have been detected, none of them can be clearly associated with the neutrino trigger. These uncatalogued sources were not bright enough to exceed the 3~$\sigma$ RASS upper limit. As the XRT failed to observe an X-ray counterpart to a neutrino trigger, this could indicate that the observation was made too late or that the source was outside the field of view of the XRT, or that the neutrino trigger was not originated by an X-ray source. Upper limits on the flux density from a potential X-ray counterpart can be derived. These limits correspond to the sensitivity reached for each 4-tile observation (see table~\ref{tab:table_alerts_swift}).
	
	\begin{landscape}	
	\begin{table}
		\centering
		\caption{Details of the 7 ANTARES triggers observed by the \textit{Swift}-XRT since 2013.}
		\label{tab:table_alerts_swift}
		\begin{threeparttable}[b]
		\begin{footnotesize}
		\begin{tabular}{cccccccc}
		\hline
		\hline
		\noalign{\smallskip}
		Trigger name &   Median error radius   &  Delay\tnote{a}  & Mean exposure  & Sensitivity & 
		 New sources (total)\tnote{b} & Counterpart & $P_{\text{reject}}^{GRB,\nu}$\tnote{c} \\
		(ANTyymmddA) &  ($^\circ$)  &  (hours)  &  (ks)  &  ($\times 10^{-13}$ erg cm$^{-2}$ s$^{-1}$)  & 
		 & candidates &  \\
		\noalign{\smallskip}
		\hline
		\noalign{\smallskip}
		ANT130722A	 & 	0.4	  &  1.1  &  1.8  & 2.74  &  4 (5) &  0  &  0.71   \\ 
		ANT130915A 	 & 	0.3	  &  6.5  &  1.4  & 3.48  &  2 (2) &  0  &  0.60   \\ 
		ANT130927A   & 	0.4   &  5.1  &  1.3  & 3.84  &  0 (1) &  0  &  0.60   \\ 
		ANT140123A   & 	0.35  &  4.7  &  0.8  & 5.99  &  1 (1) &  0  &  0.55   \\ 
		ANT140311A   & 	0.35  &  2.8  &  1.7  & 2.88  &  3 (3) &  0  &  0.68   \\ 
		ANT141220A   & 	0.4	  &  3.5  &  1.9  & 2.63  &  4 (4) &  0  &  0.67   \\ 
		ANT150129A   & 	0.35  &  1.7  &  1.9  & 2.67  &  6 (6) &  0  &  0.69   \\ 
		\noalign{\smallskip}
		\hline
		\end{tabular}
		\begin{tablenotes}
		\item[a] Delay between the neutrino trigger and the first observation by the \textit{Swift}-XRT.
		\item[b] Number of uncatalogued sources among the total number of detected sources in each 4-tile observation. Details can be found in table~\ref{tab:table_sources_swift}.
		\item[c] Probability to reject the hypothesis that the neutrino comes from a GRB.
		\end{tablenotes}
		\end{footnotesize}
		\end{threeparttable}
	\end{table}
	\end{landscape}
	
\begin{landscape} 
\begin{table}
		\centering
		\caption{Sources detected in the XRT follow-up of the ANTARES triggers.}
		\label{tab:table_sources_swift}
		\begin{threeparttable}[b]
		\begin{footnotesize}
		\begin{tabular}{cccccccc}
		\hline
		\hline
		\noalign{\smallskip}
		Alert name & Source \# & Location & Flag\tnote{a} & Exposure & Mean flux & N$_{seren}$\tnote{b} & Catalogued? \\ 
		(ANTyymmddA) & & (J2000) & & (ks) & ($\times 10^{-13}$ erg cm$^{-2}$ s$^{-1}$) &  &  \\
		\noalign{\smallskip}
		\hline
		\noalign{\smallskip}
		ANT130722A  &   1   &    04$^h$ 57$^m$ 51$^s$9~~+03$^{\circ}$ 32$'$ 01$''$    &  G  &    3.0    &   6.9 ($\pm$1.4)  &  0.2  & 1RXS J045752.5+033200\\
					&   2   &    04$^h$ 59$^m$ 46$^s$4~~+03$^{\circ}$ 23$'$ 34$''$    &  G  &    1.5    &   6.2 ($\pm$1.6)  &  0.4  & No \\
					&	3	& 	 04$^h$ 58$^m$ 06$^s$4~~+03$^{\circ}$ 27$'$ 20$''$	  &  G  &    2.6    &   2.2 ($\pm$0.6)  &  1.0  & No \\
					&	4 	& 	 04$^h$ 58$^m$ 58$^s$6~~+03$^{\circ}$ 14$'$ 54$''$ 	  &  G 	& 	 2.8 	& 	2.2 ($\pm$0.6) 	&  1.3 	& No \\
					& 	5 	& 	 04$^h$ 57$^m$ 44$^s$5~~+03$^{\circ}$ 16$'$ 33$''$ 	  &  G  &    2.7    &   0.7 ($\pm$0.4)  &  0.9  & No \\
		ANT130915A 	&   1 	&    20$^h$ 46$^m$ 05$^s$3~~-51$^{\circ}$ 49$'$ 41$''$ 	  &  R 	& 	 1.2 	& 	1.9 ($\pm$0.7) 	&  1.0  & No \\
					& 	2 	& 	 20$^h$ 44$^m$ 47$^s$6~~-51$^{\circ}$ 37$'$ 16$''$ 	  &  G  &    1.7    &   1.7 ($\pm$0.7)  &  1.0  & No \\
		ANT130927A  & 	1 	& 	 07$^h$ 08$^m$ 10$^s$8~~-68$^{\circ}$ 50$'$ 55$''$    &  G  & 	 1.5    & 	2.7 ($\pm$0.8)  &  0.6  & 1RXS J070810.2-685054\\
					&   2   &    07$^h$ 06$^m$ 06$^s$3~~-68$^{\circ}$ 35$'$ 57$''$    &  G  &    1.6    &   1.6 ($\pm$0.6)  &  1.3  & No \\
					&   3   &    07$^h$ 09$^m$ 32$^s$8~~-68$^{\circ}$ 35$'$ 31$''$    &  G  &    1.9    &   1.1 ($\pm$0.5)  &  1.7  & No \\
					&   4   &    07$^h$ 05$^m$ 44$^s$5~~-68$^{\circ}$ 46$'$ 25$''$    &  R  &    2.0    &   1.2 ($\pm$0.6)  &  1.7  & No \\
		ANT140123A  & 	1 	& 	 07$^h$ 00$^m$ 20$^s$4~~+00$^{\circ}$ 49$'$ 27$''$    &  G  &    1.4    & 	3.3 ($\pm$1.1)  &  0.6  & No \\
		ANT140311A  & 	1 	&    15$^h$ 50$^m$ 10$^s$9~~+31$^{\circ}$ 39$'$ 30$''$    &  G  &    3.0    & 	2.6 ($\pm$1.2)  &  1.2  & No \\
					& 	2 	&    15$^h$ 49$^m$ 31$^s$2~~+31$^{\circ}$ 29$'$ 14$''$    &  G  &    1.5    & 	1.7 ($\pm$0.8)  &  0.8  & No \\
					& 	3 	&    15$^h$ 51$^m$ 02$^s$4~~+31$^{\circ}$ 56$'$ 15$''$    &  G  &    1.4    & 	2.9 ($\pm$1.2)  &  1.1  & No \\
		ANT141220A  & 	1 	&    04$^h$ 47$^m$ 35$^s$2~~-46$^{\circ}$ 10$'$ 54$''$    &  G  &    1.5    & 	1.5 ($\pm$0.7)  &  1.3  & No \\
					& 	2 	&    04$^h$ 46$^m$ 10$^s$9~~-45$^{\circ}$ 58$'$ 44$''$    &  G  &    1.7    & 	1.4 ($\pm$0.7)  &  1.9  & No \\
		ANT150129A  & 	1 	&    02$^h$ 19$^m$ 31$^s$6~~+04$^{\circ}$ 51$'$ 18$''$    &  G  &    1.9    & 	1.3 ($\pm$0.6)  &  1.4  & No \\
					& 	2 	&    02$^h$ 19$^m$ 16$^s$2~~+04$^{\circ}$ 34$'$ 29$''$    &  G  &    1.4    & 	0.3 ($\pm$0.2)  &  1.5  & No \\
					& 	3 	&    02$^h$ 17$^m$ 52$^s$6~~+04$^{\circ}$ 49$'$ 32$''$ 	  &  G  &    1.4    & 	1.4 ($\pm$0.6)  &  1.4  & No \\

		\noalign{\smallskip}
		\hline

		\end{tabular}
		\begin{tablenotes}
		\item[a] Indicates how likely the source is to be real. G=Good, R=Reasonable, P=Poor.
		\item[b] Number of expected serendipitous sources in the field, at least as bright as the detected object.
		\end{tablenotes}
		\end{footnotesize}
		\end{threeparttable}
	\end{table}
	\end{landscape}


\section{Limit on a GRB association}

GRBs are the major candidate sources of high-energy neutrinos among the population of fast transient sources. In the standard fireball model of GRB~\citep{Kumar2014}, internal shocks between faster and slower shells of plasma expulsed in a highly relativistic jet would be responsible for the observed prompt gamma-ray emission, resulting in photohadronic interactions. Hadronic models of GRBs~\citep{Waxman1997,Guetta2001} suggest that there could be a significant proton loading in GRB's jets. If these protons are sufficiently accelerated they can interact with the ambient gamma photon field via p$\gamma$ mechanisms~\citep{Hummer2010,Meszaros2012}. The photohadronic interactions would produce charged pions, subsequently decaying into high-energy neutrinos. Details of the physical processes taking place in the GRB prompt emission are still under debate and the neutrino expectations are highly dependent on the GRB prompt model used. 

According to the standard framework of internal shocks, the expected number of detectable neutrinos from two luminous bursts GRB110918A, GRB130427A and one typical 
burst GRB081008 can be estimated: see~\cite{ANTARES_GRB2013} for details on the calculation method. The presented results use the predicted neutrino spectra from the NeuCosmA model~\citep[][]{Hummer2010,Hummer2012} that simulates the complete p$\gamma$ mechanism ($\Delta$-resonance, kaon production, multiple pions decay) in the internal shock GRB model. Standard values for the unknown physical parameters are assumed such as the Lorentz factor $\Gamma = 316$, the baryonic loading $f_p = 10$ and the fraction of internal energy given to the magnetic field and electrons $\epsilon_B = \epsilon_e = 0.1$. The expected number of prompt neutrinos for each individual GRB is $\mu_s =3.4\times10^{-2}$, $\mu_s = 6.6\times10^{-3}$ and $\mu_s = 2.3\times10^{-5}$ for GRB110918A, GRB130427A and GRB081008, respectively. So far, no neutrino signal has been observed in coincidence with a GRB. Nevertheless, the probability to observe $n$ neutrinos with a given signal $\mu_s$ can be estimated with a Poissonian distribution. The probabilities to detect at least one neutrino from the mentioned GRBs are $P(X\geq1 | 3.4\times10^{-2})\sim 3.4$~\%, $P(X\geq1 | 6.6\times10^{-3})\sim 0.7$~\% and $P(X\geq1 | 2.3\times10^{-5})\sim 0.002$~\%, respectively. Albeit these discovery probabilities are weak, one should not forget that they are affected by strong theoretical uncertainties.

The TAToO program offers an opportunity to test the GRB origin for each detected neutrino with the advantage of not depending on an underlying GRB model. The probability, $P_s$, to detect a serendipitous GRB afterglow lying in the follow-up region for each neutrino alert is given by:
\begin{equation}
P_s(\text{GRB},\nu) = R_{\text{GRB}} \times p(\text{ag}|\text{GRB}) \times \frac{\Omega}{4\pi} \times T_{\text{obs}}
\end{equation}
where $R_{\text{GRB}}$ is the average GRB discovery rate (1000/yr) in the entire sky \citep{2009ARA&A..47..567G} and $p(\text{ag}|\text{GRB})$ is the probability of detecting the afterglow counterpart of a detected GRB. This probability is 0.4 and 0.5 for TAROT and ROTSE respectively~\citep{Klotz_0.4_in_prep,2005ApJ...631.1032R}, and 0.95 for \textit{Swift}~\citep{2009ARA&A..47..567G}. The solid angle, $\Omega$, viewed by the telescopes is given by $\Omega = 2\pi(1-cos(\theta))$ with $\theta \sim 1^{\circ}$ for TAROT and ROTSE, and $\theta\sim 0.2^{\circ}$ for the \textit{Swift}-XRT. $T_{\text{obs}}$ is the exposure time of the observation of each neutrino alert. The optical follow-up lasts about 20 minutes, while the X-ray follow-up lasts 2 ks for each tile.
Thus, $P_s(\text{GRB},\nu) = 1.2 \times10^{-6}$ and $1.5\times10^{-6}$ for each alert with an optical follow-up with TAROT and ROTSE respectively, and $P_s(\text{GRB},\nu) = 1.8\times10^{-7}$ for each alert with X-ray images. Consequently, a discovery of a GRB afterglow would be unambiguously associated with the neutrino alert.

\subsection{Optical GRB association}
\label{sec:optical_grb_assoc}

\begin{figure}[!ht]
		\centering
		\includegraphics[width=0.65\textwidth]{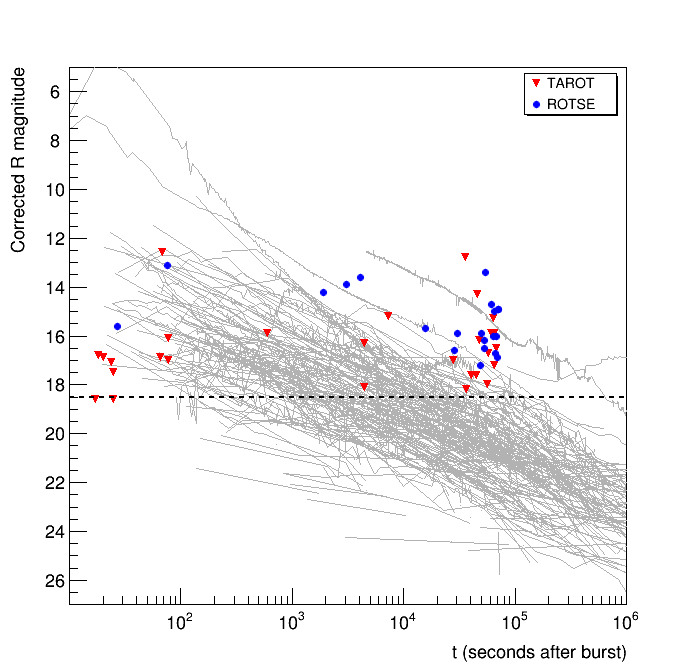}
		\caption{Grey lines: Corrected R magnitude as a function of time for 301 GRB afterglows observed from 1997 to 2014 by optical telescopes. Red \& blue dots: upper limits on GRB magnitudes for neutrino alerts observed by TAROT and ROTSE respectively. Each point represents the first image of the observation, corresponding to an exposure of 180 seconds for TAROT images, and 20 or 60 seconds for ROTSE images. The horizontal dashed line corresponds to the maximum sensitivity of the telescopes.}
		\label{fig:GRB_limits_optical}
	\end{figure}

Because in this study no optical counterpart has been observed in coincidence with the 42 neutrino alerts, the probability to reject a GRB associated with each neutrino alert can be directly estimated. To do so, a comparison is made between the optical upper limits obtained for each neutrino alert with optical 
afterglow light curves of GRBs detected from 1997 to 2014, as shown in figure \ref{fig:GRB_limits_optical}.

\begin{figure}[!ht]
		\centering
		\includegraphics[trim={0 6.5cm 0 5cm},clip,width=0.65\textwidth]{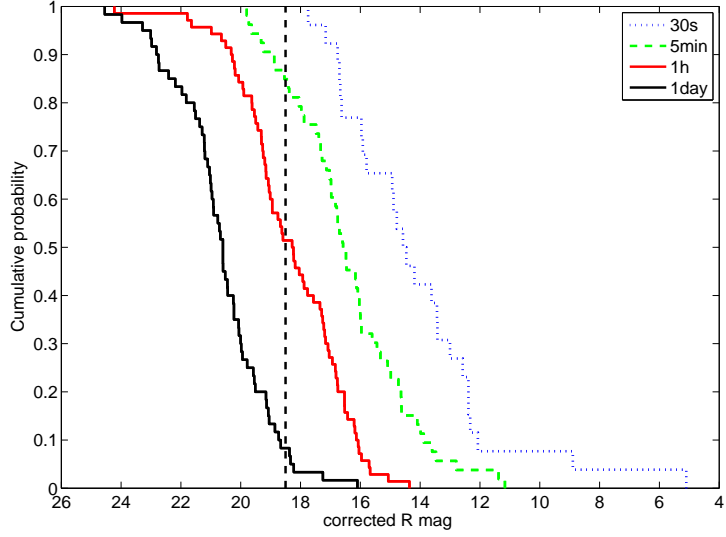}
		\caption{Cumulative distribution of afterglow magnitudes for 301 detected GRBs (figure~\ref{fig:GRB_limits_optical}). Each line corresponds to different times after burst. The vertical dashed line represents the limiting magnitude of the optical telescopes.}
		\label{fig:cumul_proba_optique}
	\end{figure}

The cumulative distribution functions (CDFs) of afterglow magnitudes (corrected by the galactic extinction) are computed at times coincident with the first follow-up observation of the neutrino alerts. Figure~\ref{fig:cumul_proba_optique} shows these CDFs at typical times t = 30 seconds, 5 minutes, 1 hour and 1 day after the GRB in the observer frame. The probability, $P_{\text{reject}}^{\text{GRB},\nu}$, to reject a GRB origin hypothesis for each detected neutrino can be directly extracted from the CDFs, accounting for the field of view of telescopes\footnote{The probabilities extracted from the CDFs are multiplied by the percentage of the ANTARES PSF covered by the telescopes (see table~\ref{tab:perf_tatoo}).}. The $P_{\text{reject}}^{\text{GRB},\nu}$ are listed in table~\ref{tab:table_optical_alerts} for each alert.

According to figure~\ref{fig:GRB_limits_optical}, stringent upper limits on the early afterglow magnitude a few minutes after the neutrino trigger can be set. For the two neutrino alerts ANT121010A and ANT150122A, a GRB origin is strongly rejected since no early afterglow at a time t $<$ 30 seconds after the burst has ever been observed at weaker magnitude than $\sim 18$. Except for the alert ANT121206A (low limiting magnitude), a GRB association is rejected at a better than 80~\% confidence level within 1 minute after the neutrino trigger. On the contrary, when the optical follow-up starts more than a few minutes after the trigger, due to the sensitivity of the telescopes, the derived upper limits do not constraint a GRB origin for the detected neutrinos. 

\subsection{X-ray GRB association}

\begin{figure}[!ht]
		\centering
		\includegraphics[width=0.7\textwidth]{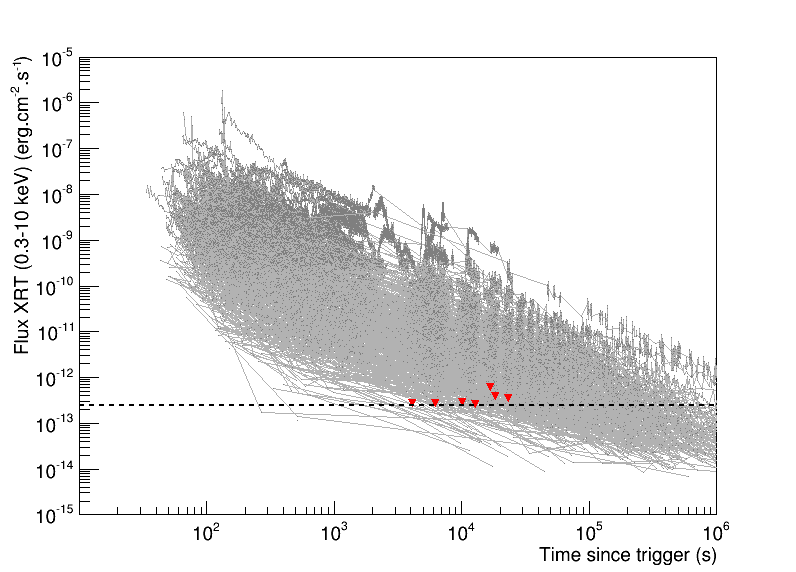}
		\caption{Grey lines: 689 X-ray afterglow fluxes in the energy band from 0.3 to 10 keV detected by the \textit{Swift}-XRT from 2007 to 2015 as a function of 
		time. The upper limits on GRB fluxes for 7 neutrino alerts are represented by red triangles. The horizontal dashed line corresponds to the sensitivity 
		reached with a 2 ks exposure.}
		\label{fig:GRB_limits_XRT}
	\end{figure}

As no X-ray counterpart has been identified in coincidence with the 7 neutrino alerts, an analysis similar to that in the optical domain is performed with the 689 X-ray afterglow light curves detected by \textit{Swift}-XRT from 2007 to 2015 (figure~\ref{fig:GRB_limits_XRT}). A comparison is made between these light curves and the X-ray upper limits obtained for the 7 neutrino alerts.
	
\begin{figure}[!h]
		\centering
		\includegraphics[trim={0 0 2cm 14cm},clip,width=0.65\textwidth]{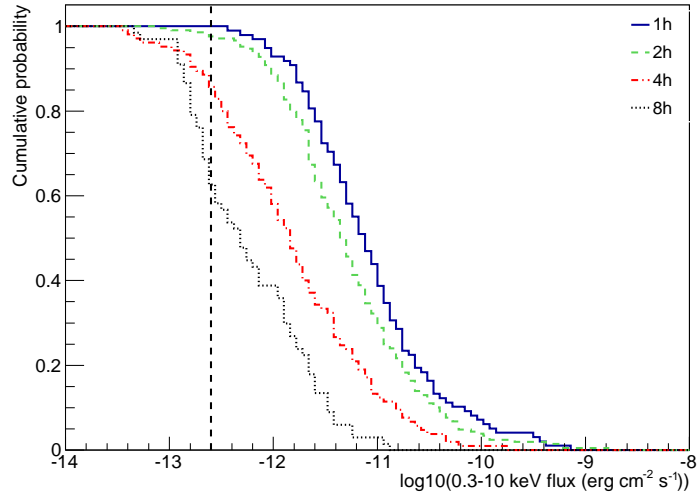}
		\caption{Cumulative distribution of X-ray afterglow magnitudes for 689 GRBs detected by the \textit{Swift}-XRT 
		since 2007. Each line represents different times after bursts. The vertical dashed line represents the sensitivity 
		reached with a 2 ks exposure.}
		\label{fig:cumul_proba_X}
	\end{figure}

The CDFs of the X-ray afterglow fluxes are also computed at times coincident with the first XRT observation of neutrino alerts. Figure~\ref{fig:cumul_proba_X} shows these CDFs at t = 1, 2, 4 and 8 hours after the GRB in the observer frame. The probabilities to reject a GRB origin of the neutrino events are calculated as previously\footnote{For the X-ray follow-up, about 72~\% of the ANTARES PSF is covered.} and are listed in table~\ref{tab:table_alerts_swift}. According to figure~\ref{fig:GRB_limits_XRT}, if each detected neutrino comes from a GRB, strong upper limits on the GRB afterglow X-ray flux can be set, $F_X\lesssim 2.5\times10^{-13}$ erg cm$^{-2}$ s$^{-1}$. For the alert ANT130722A, the probability to reject a GRB origin is as high as 71~\% with an upper limit on the X-ray flux of $\sim 2.74\times10^{-13}$ erg cm$^{-2}$ s$^{-1}$ only 1.1 hours after the neutrino trigger. Clearly, a GRB origin is very unlikely for these neutrino events, since few GRBs are known to have weaker X-ray afterglows than these upper limits. 

\section{Improvements for TAToO}

The optical and X-ray upper limits derived from the follow-up of 48 neutrino alerts have proved that the TAToO system is able to quickly check the GRB origin 
of the high-energy neutrinos detected by ANTARES. Further improvements, such as starting the follow-up as soon as possible after the trigger, would enable more definite conclusions on the origin of these high-energy neutrinos.

\subsection{Optical domain}
While small robotic telescopes, such as TAROT and ROTSE, are well-suited to quickly respond to ANTARES alerts, their diameters do not permit magnitudes deeper than $\text{R} \sim 18.5$ to be reached. This limits the ability of the TAToO program to investigate a GRB origin of the detected neutrinos to within only a few minutes after the ANTARES trigger. A one-metre-class robotic telescope is crucial to efficiently extend the optical follow-up of ANTARES alerts with stronger 
constraints on the neutrino/GRB association up to a few hours after the neutrino trigger. 

Since 2013, the Zadko telescope has been included in the optical telescope network~\citep{Coward2010}. The Zadko telescope is a one metre fully robotic telescope located at the Gingin observatory in Western Australia, covering a field of view of about 0.15 square degrees. This telescope is very interesting for the TAToO program as it is located in the area where most of the ANTARES neutrino alerts are immediately visible (figure~\ref{fig:map_tarot_rotse}). Thanks to its larger aperture, about 3 times larger than the TAROT and ROTSE telescopes, the limiting magnitude of the Zadko telescope with only 60 seconds of exposure is 1.4 magnitudes deeper compared to the TAROT telescopes, with 180 seconds of exposure. This represents a gain of about a factor of 4 on the optical sensitivity. Due to its limited field of view, observations are composed of 7 tiles of 60 seconds exposure each, covering about 85~\% of the neutrino alert error box. The best response of Zadko to an ANTARES alert achieved so far is about 70 seconds after the neutrino trigger. 
The combination of its reasonable response, its larger aperture and its good location will help to better constrain the origin of high-energy neutrinos. 

In addition, five telescopes from the MASTER network~\citep[][]{Lipunov2010,Lipunov2012} have also joined the TAToO program since March 2015. They consist of pairs of fully robotic telescopes with a diameter of 0.40 m covering a field of view of 8 square degrees per pair of telescopes. They are located in Russia and South Africa, and significantly increase the coverage of the TAToO telescope network. The first alert sendings with MASTER demonstrate the rapid response time to the TAToO alert of about 24 seconds after the neutrino trigger, which is roughly the same performance as TAROT. Thanks to this new partnership with MASTER, the very early optical follow-up (less than a minute) of ANTARES alerts will be considerably more efficient at constraining the origin of each neutrino.

Finally, the addition of an infrared follow-up to TAToO alerts would be very helpful, since this part of the electromagnetic spectrum is less subject to dust absorption. In the case of GRBs, this new wavelength domain will allow a larger population of GRBs (e.g. extinguished GRB afterglows) to be explored, with a correspondingly higher neutrino detection potential. 

\subsection{X-ray domain}
In the X-ray domain, the fastest response to the 7 neutrino alerts was about 1 hour. While strong constraints on the GRB origin of the high-energy neutrinos 
detected are set, a faster response or a better X-ray sensitivity would almost undoubtedly allow for the rejection (or confirmation) of the GRB origin hypothesis. 

With the next generation of neutrino telescopes, such as KM3NeT (see below), the X-ray follow-up will be possible with the arrival of new high-energy missions. In the case of the SVOM mission ($\sim 2021$) ~\citep{SVOM}, Target-of-Opportunity alerts will be accepted, and X-ray follow-up of neutrino alerts by the MXT instrument~\citep{MXT_SVOM} is under discussion. The sensitivity of the SVOM-MXT ($3\times10^{-12}$ erg cm$^{-2}$ s$^{-1}$ with 1 ks exposure) would allow roughly the same constraints as for the \textit{Swift}-XRT to be reached. Moreover, with the field of view of the MXT ($\sim 1^{\circ}\times1^{\circ}$), only one tile will be needed, allowing time to be saved for other neutrino alert follow-ups.

\subsection{The future neutrino detector KM3NeT}

KM3NeT\footnote{www.km3net.org} is the next-generation of neutrino telescope based on the same detection techniques as ANTARES in deep seawater. The size of the KM3NeT detector, at several cubic kilometers, will increase the sensitivity to neutrino fluxes by about a factor 50 compared to ANTARES. By convolving the expected KM3NeT response function with the neutrino flux expected from the NeuCosmA model, the number of high-energy neutrinos, $\mu_s$, expected from GRB110918A and GRB130427A can be estimated. For these two GRBs, $\mu_s = 1$ and $\mu_s = 0.4$, respectively, assuming the muon channel only. This expectation is clearly encouraging, and the detection of high-energy neutrinos in coincidence with GRBs will be achievable with the future KM3NeT detector. Including the faster response of the future X-ray telescopes and a better optical/X-ray sensitivity of the TAToO telescope network, a factor of $\sim 100$ in sensitivity would be gained for probing a GRB origin for high-energy neutrinos.

\section{Conclusion}	
Optical and X-ray follow-ups of the ANTARES neutrino alerts have been running stably since 2010 and mid 2013 respectively. Beginning 2015, 150 and 7 alerts have been sent to the robotic telescopes TAROT and ROTSE, and to the XRT telescope on board of the \textit{Swift} satellite. The main advantage of the TAToO program is that it is able to send alerts within a few seconds after the neutrino detection with a precision better than 0.5$^{\circ}$ for high-energy neutrinos. Early follow-up has been performed for 42 alerts with delays as low as 17 seconds between the neutrino detection and the start of the acquisition of the first image. 
The image analysis has not yet discovered any transient sources associated with the selected high-energy neutrinos. Upper limits on the magnitude of possible transient sources have been derived. Compared to detected GRB afterglow light curves, the very rapid response time has allowed stringent constraints on the GRB origin of individual neutrinos to be placed. Even if the response time of the XRT follow-up is larger, the early observations of the neutrino alerts have allowed the GRB origin for these 7 neutrinos to be excluded with a high confidence level. 


\acknowledgments

The authors acknowledge the financial support of the funding agencies:
Centre National de la Recherche Scientifique (CNRS), Commissariat \`a
l'\'energie atomique et aux \'energies alternatives (CEA), 
Commission Europ\'eenne (FEDER fund
and Marie Curie Program), R\'egion Ile-de-France (DIM-ACAV), R\'egion Alsace (contrat CPER), R\'egion
Provence-Alpes-C\^ote d'Azur, D\'epartement du Var and Ville de La
Seyne-sur-Mer, France; Bundesministerium f\"ur Bildung und Forschung
(BMBF), Germany; Istituto Nazionale di Fisica Nucleare (INFN), Italy;
Stichting voor Fundamenteel Onderzoek der Materie (FOM), Nederlandse
organisatie voor Wetenschappelijk Onderzoek (NWO), the Netherlands;
Council of the President of the Russian Federation for young
scientists and leading scientific schools supporting grants, Russia;
National Authority for Scientific Research (ANCS), Romania; Ministerio
de Ciencia e Innovaci\'on (MICINN), Prometeo of Generalitat Valenciana
and MultiDark, Spain; Agence de l'Oriental and CNRST, Morocco. We also
acknowledge the technical support of Ifremer, AIM and Foselev Marine
for the sea operation and the CC-IN2P3 for the computing facilities. We 
gratefully acknowledge financial support from the OCEVU LabEx, France. 

\bibliographystyle{unsrt} 
\bibliography{biblio.bib}

\begin{thebibliography}{10}

\bibitem{Chiarusi}
T.~{Chiarusi} and M.~{Spurio}.
\newblock {High-energy astrophysics with neutrino telescopes}.
\newblock {\em European Physical Journal C}, 65:649--701, February 2010.

\bibitem{Anchordoqui}
L.~A. {Anchordoqui} and T.~{Montaruli}.
\newblock {In Search of Extraterrestrial High-Energy Neutrinos}.
\newblock {\em Annual Review of Nuclear and Particle Science}, 60:129--162,
  November 2010.

\bibitem{Kouveliotou}
C.~{Kouveliotou et al.}
\newblock {Identification of two classes of gamma-ray bursts}.
\newblock {\em ApJ}, 413:L101--L104, August 1993.

\bibitem{FermiBlazars}
A.~A. {Abdo et al.}
\newblock {Gamma-ray Light Curves and Variability of Bright Fermi-detected
  Blazars}.
\newblock {\em ApJ}, 722:520--542, October 2010.

\bibitem{CCSNe_variability}
A.~{Burrows}.
\newblock {Colloquium: Perspectives on core-collapse supernova theory}.
\newblock {\em Reviews of Modern Physics}, 85:245--261, January 2013.

\bibitem{ANTARES}
M.~{Ageron et al.}
\newblock {ANTARES: The first undersea neutrino telescope}.
\newblock {\em Nuclear Instruments and Methods in Physics Research A},
  656:11--38, November 2011.

\bibitem{IceCube}
A.~{Achterberg et al.}
\newblock {First year performance of the IceCube neutrino telescope}.
\newblock {\em Astroparticle Physics}, 26:155--173, October 2006.

\bibitem{IC_paper_1}
M.~G. {Aartsen et al.}
\newblock {Evidence for High-Energy Extraterrestrial Neutrinos at the IceCube
  Detector}.
\newblock {\em Science}, 342:1, November 2013.

\bibitem{IC_paper_2}
M.~G. {Aartsen et al.}
\newblock {Observation of High-Energy Astrophysical Neutrinos in Three Years of
  IceCube Data}.
\newblock {\em Phys. Rev. Lett.}, 113(10):101101, September 2014.

\bibitem{GRB_ANTARES}
S.~{Adri{\'a}n-Mart{\'{\i}}nez et al.}
\newblock {Search for muon neutrinos from gamma-ray bursts with the ANTARES
  neutrino telescope using 2008 to 2011 data}.
\newblock {\em A\&A}, 559:A9, November 2013.

\bibitem{AGN_ANTARES}
S.~{Adri{\'a}n-Mart{\'{\i}}nez et al.}
\newblock {Search for neutrino emission from gamma-ray flaring blazars with the
  ANTARES telescope}.
\newblock {\em Astroparticle Physics}, 36:204--210, August 2012.

\bibitem{microquasars_ANTARES}
S.~{Adri{\'a}n-Mart{\'{\i}}nez et al.}
\newblock {A search for time dependent neutrino emission from microquasars with
  the ANTARES telescope}.
\newblock {\em Journal of High Energy Astrophysics}, 3:9--17, September 2014.

\bibitem{TAToO}
M.~{Ageron et al.}
\newblock {The ANTARES telescope neutrino alert system}.
\newblock {\em Astroparticle Physics}, 35:530--536, March 2012.

\bibitem{TAROT_klotz}
A.~{Klotz}, M.~{Bo{\"e}r}, J.~L. {Atteia}, and B.~{Gendre}.
\newblock {Early Optical Observations of Gamma-Ray Bursts by the TAROT
  Telescopes: Period 2001-2008}.
\newblock {\em AJ}, 137:4100--4108, May 2009.

\bibitem{ROTSE_akerlof}
C.~W. {Akerlof et al.}
\newblock {The ROTSE-III Robotic Telescope System}.
\newblock {\em PASP}, 115:132--140, January 2003.

\bibitem{XRT}
D.~N. {Burrows et al.}
\newblock {The Swift X-Ray Telescope}.
\newblock {\em Space Sci. Rev.}, 120:165--195, October 2005.

\bibitem{IceCube_follow-up}
R.~{Abbasi et al.}
\newblock {Searching for soft relativistic jets in core-collapse supernovae
  with the IceCube optical follow-up program}.
\newblock {\em A\&A}, 539:A60, March 2012.

\bibitem{point_source_ANTARES}
S.~{Adri{\'a}n-Mart{\'{\i}}nez et al.}
\newblock {Searches for Point-like and Extended Neutrino Sources Close to the
  Galactic Center Using the ANTARES Neutrino Telescope}.
\newblock {\em ApJ}, 786:L5, May 2014.

\bibitem{GRBs1}
E.~{Waxman} and J.~{Bahcall}.
\newblock {High Energy Neutrinos from Cosmological Gamma-Ray Burst Fireballs}.
\newblock {\em Phys. Rev. Lett.}, 78:2292--2295, March 1997.

\bibitem{GRBs2}
P.~{M{\'e}sz{\'a}ros} and E.~{Waxman}.
\newblock {TeV Neutrinos from Successful and Choked Gamma-Ray Bursts}.
\newblock {\em Phys. Rev. Lett.}, 87(17):171102, October 2001.

\bibitem{GRBs3}
C.~D. {Dermer} and A.~{Atoyan}.
\newblock {High-Energy Neutrinos from Gamma Ray Bursts}.
\newblock {\em Phys. Rev. Lett.}, 91(7):071102, August 2003.

\bibitem{GRBs4}
S.~{Razzaque}, P.~{M{\'e}sz{\'a}ros}, and E.~{Waxman}.
\newblock {High Energy Neutrinos from Gamma-Ray Bursts with Precursor
  Supernovae}.
\newblock {\em Phys. Rev. Lett.}, 90(24):241103, June 2003.

\bibitem{ccSNe}
S.~{Ando} and J.~F. {Beacom}.
\newblock {Revealing the Supernova Gamma-Ray Burst Connection with TeV
  Neutrinos}.
\newblock {\em Phys. Rev. Lett.}, 95(6):061103, August 2005.

\bibitem{AGN_hadron_model}
M.~{B{\"o}ttcher}, A.~{Reimer}, K.~{Sweeney}, and A.~{Prakash}.
\newblock {Leptonic and Hadronic Modeling of Fermi-detected Blazars}.
\newblock {\em ApJ}, 768:54, May 2013.

\bibitem{BBFit_ANTARES}
J.~A. {Aguilar et al.}
\newblock {A fast algorithm for muon track reconstruction and its application
  to the ANTARES neutrino telescope}.
\newblock {\em Astroparticle Physics}, 34:652--662, April 2011.

\bibitem{AAFit_ANTARES}
S.~{Adri{\'a}n-Mart{\'{\i}}nez et al.}
\newblock {First Search for Point Sources of High-energy Cosmic Neutrinos with
  the ANTARES Neutrino Telescope}.
\newblock {\em ApJ}, 743:L14, December 2011.

\bibitem{GWGC}
D.~J. {White}, E.~J. {Daw}, and V.~S. {Dhillon}.
\newblock {A list of galaxies for gravitational wave searches}.
\newblock {\em Classical and Quantum Gravity}, 28(8):085016, April 2011.

\bibitem{GCN_barthelmy}
S.~D. {Barthelmy et al.}
\newblock {The GRB coordinates network (GCN): A status report}.
\newblock In C.~A. {Meegan}, R.~D. {Preece}, and T.~M. {Koshut}, editors, {\em
  Gamma-Ray Bursts, 4th Hunstville Symposium}, volume 428 of {\em American
  Institute of Physics Conference Series}, pages 99--103, May 1998.

\bibitem{SExtractor}
E.~{Bertin} and S.~{Arnouts}.
\newblock {SExtractor: Software for source extraction.}
\newblock {\em A\&AS, Supplement}, 117:393--404, June 1996.

\bibitem{SCAMP}
E.~{Bertin}.
\newblock {Automatic Astrometric and Photometric Calibration with SCAMP}.
\newblock In C.~{Gabriel}, C.~{Arviset}, D.~{Ponz}, and S.~{Enrique}, editors,
  {\em Astronomical Data Analysis Software and Systems XV}, volume 351 of {\em
  Astronomical Society of the Pacific Conference Series}, page 112, July 2006.

\bibitem{LePHARE}
S.~{Arnouts} and O.~{Ilbert}.
\newblock {LePHARE: Photometric Analysis for Redshift Estimate}.
\newblock Astrophysics Source Code Library, August 2011.

\bibitem{NOMAD}
N.~{Zacharias et al.}
\newblock {The Naval Observatory Merged Astrometric Dataset (NOMAD)}.
\newblock In {\em American Astronomical Society Meeting Abstracts}, volume~36
  of {\em Bulletin of the American Astronomical Society}, page 1418, December
  2004.

\bibitem{IRAF}
D.~{Tody}.
\newblock {The IRAF Data Reduction and Analysis System}.
\newblock In D.~L. {Crawford}, editor, {\em Instrumentation in astronomy VI},
  volume 627 of {\em SPIE Conference Series}, page 733, January 1986.

\bibitem{Schlegel}
D.~J. {Schlegel}, D.~P. {Finkbeiner}, and M.~{Davis}.
\newblock {Maps of Dust Infrared Emission for Use in Estimation of Reddening
  and Cosmic Microwave Background Radiation Foregrounds}.
\newblock {\em ApJ}, 500:525--553, June 1998.

\bibitem{Swift}
N.~{Gehrels et al.}
\newblock {The Swift Gamma-Ray Burst Mission}.
\newblock {\em ApJ}, 611:1005--1020, August 2004.

\bibitem{Evans_2014}
P.~A. {Evans et al.}
\newblock {1SXPS: A Deep Swift X-Ray Telescope Point Source Catalog with Light
  Curves and Spectra}.
\newblock {\em ApJS}, 210:8, January 2014.

\bibitem{Moretti}
A.~{Moretti et al.}
\newblock {The swift-XRT imaging performances and serendipitous survey}.
\newblock In {\em SPIE Conference Series}, volume 6688, page~0, September 2007.

\bibitem{Goad}
M.~R. {Goad et al.}
\newblock {Accurate early positions for Swift GRBs: enhancing X-ray positions
  with UVOT astrometry}.
\newblock {\em A\&A}, 476:1401--1409, December 2007.

\bibitem{Evans_2009}
P.~A. {Evans}, A.~P. {Beardmore}, J.~P. {Osborne}, and G.~A. {Wynn}.
\newblock {The unusual 2006 dwarf nova outburst of GK Persei}.
\newblock {\em MNRAS}, 399:1167--1174, November 2009.

\bibitem{Evans_2007}
P.~A. {Evans et al.}
\newblock {An online repository of Swift/XRT light curves of {$\gamma$}-ray
  bursts}.
\newblock {\em A\&A}, 469:379--385, July 2007.

\bibitem{RASS}
W.~{Voges et al.}
\newblock {The ROSAT all-sky survey bright source catalogue}.
\newblock {\em A\&A}, 349:389--405, September 1999.

\bibitem{Evans_IceCube_2015}
P.~A. {Evans et al.}
\newblock {Swift follow-up of IceCube triggers, and implications for the
  Advanced-LIGO era}.
\newblock {\em ArXiv e-prints}, January 2015.

\bibitem{Kumar2014}
P.~{Kumar} and B.~{Zhang}.
\newblock {The Physics of Gamma-Ray Bursts and Relativistic Jets}.
\newblock {\em ArXiv e-prints}, October 2014.

\bibitem{Waxman1997}
E.~{Waxman} and J.~{Bahcall}.
\newblock {High Energy Neutrinos from Cosmological Gamma-Ray Burst Fireballs}.
\newblock {\em Phys. Rev. Lett.}, 78:2292--2295, March 1997.

\bibitem{Guetta2001}
D.~{Guetta}, M.~{Spada}, and E.~{Waxman}.
\newblock {On the Neutrino Flux from Gamma-Ray Bursts}.
\newblock {\em ApJ}, 559:101--109, September 2001.

\bibitem{Hummer2010}
S.~{H{\"u}mmer}, M.~{R{\"u}ger}, F.~{Spanier}, and W.~{Winter}.
\newblock {Simplified Models for Photohadronic Interactions in Cosmic
  Accelerators}.
\newblock {\em ApJ}, 721:630--652, September 2010.

\bibitem{Meszaros2012}
P.~{M{\'e}sz{\'a}ros}, K.~{Asano}, and P.~{Veres}.
\newblock {Gamma Ray Bursts: recent results and connections to very high energy
  Cosmic Rays and Neutrinos}.
\newblock {\em ArXiv e-prints}, September 2012.

\bibitem{ANTARES_GRB2013}
S.~{Adri{\'a}n-Mart{\'{\i}}nez et al.}
\newblock {Search for muon neutrinos from gamma-ray bursts with the ANTARES
  neutrino telescope using 2008 to 2011 data}.
\newblock {\em A\&A}, 559:A9, November 2013.

\bibitem{Hummer2012}
S.~{H{\"u}mmer}, P.~{Baerwald}, and W.~{Winter}.
\newblock {Neutrino Emission from Gamma-Ray Burst Fireballs, Revised}.
\newblock {\em Phys. Rev. Lett.}, 108(23):231101, June 2012.

\bibitem{2009ARA&A..47..567G}
N.~{Gehrels}, E.~{Ramirez-Ruiz}, and D.~B. {Fox}.
\newblock {Gamma-Ray Bursts in the Swift Era}.
\newblock {\em ARA\&A}, 47:567--617, September 2009.

\bibitem{Klotz_0.4_in_prep}
A.~{Klotz}.
\newblock {}.
\newblock {\em Paper in prep.}, 2015.

\bibitem{2005ApJ...631.1032R}
E.~S. {Rykoff et al.}
\newblock {A Search for Untriggered GRB Afterglows with ROTSE-III}.
\newblock {\em ApJ}, 631:1032--1038, October 2005.

\bibitem{Coward2010}
D.~M. {Coward et al.}
\newblock {The Zadko Telescope: A Southern Hemisphere Telescope for Optical
  Transient Searches, Multi-Messenger Astronomy and Education}.
\newblock {\em Publications of the Astron. Soc. of Australia}, 27:331--339,
  September 2010.

\bibitem{Lipunov2010}
V.~{Lipunov et al.}
\newblock {Master Robotic Net}.
\newblock {\em Advances in Astronomy}, 2010:30, 2010.

\bibitem{Lipunov2012}
V.~{Lipunov et al.}
\newblock {MASTER global robotic net}.
\newblock In {\em Astronomical Society of India Conference Series}, volume~7 of
  {\em Astronomical Society of India Conference Series}, page 275, 2012.

\bibitem{SVOM}
O.~{Godet et al.}
\newblock {The Chinese-French SVOM Mission: studying the brightest astronomical
  explosions}.
\newblock In {\em Society of Photo-Optical Instrumentation Engineers (SPIE)
  Conference Series}, volume 8443, page~1, September 2012.

\bibitem{MXT_SVOM}
D.~{G{\"o}tz et al.}
\newblock {The microchannel x-ray telescope for the gamma-ray burst mission
  SVOM}.
\newblock In {\em Society of Photo-Optical Instrumentation Engineers (SPIE)
  Conference Series}, volume 9144, page~23, July 2014.

\end{thebibliography}

\end{document}